\documentclass[%
aip,
amsmath,amssymb,
reprint,%
]{revtex4-1}

\usepackage{dblfloatfix}
\usepackage{graphicx}
\usepackage{amsmath}
\usepackage{mathtools}
\usepackage[titletoc, toc, page]{appendix}
\usepackage{chemformula}
\usepackage{amssymb}
\usepackage{float}
\usepackage{hyperref}
\usepackage[]{color}
\usepackage{subfig}
\usepackage{afterpage}%
 \DeclareUnicodeCharacter{2212}{\ensuremath{-}}

\begin{document}		 
	
	\author{Yan Levin}
	\email{levin@if.ufrgs.br}
	\affiliation{Instituto de F\'isica, Universidade Federal do Rio Grande do Sul, Caixa Postal 15051, CEP 91501-970, Porto Alegre, RS, Brazil.}
	
	\author{Amin Bakhshandeh}
	
	\affiliation{Instituto de F\'isica, Universidade Federal do Rio Grande do Sul, Caixa Postal 15051, CEP 91501-970, Porto Alegre, RS, Brazil.}

	\title{A new method for reactive constant pH simulations}

	\begin{abstract}
		We present a simulation method that allows us to calculate the titration curves for 
		systems undergoing protonation/deprotonation reactions -- such as charged colloidal suspensions with acidic/basic surface groups, polyelectrolytes, polyampholytes, proteins,  etc.  The new approach allows us to simultaneously obtain titration curves both for systems in contact with salt and acid reservoir (semi-grand canonical ensemble)  and for isolated suspensions (canonical ensemble).  To treat the electrostatic  interactions, we present a new method based on Ewald summation  -- which accounts for the existence of both Bethe and Donnan potentials within the simulation cell.   We show  that the Donnan potential affects dramatically the pH of suspension.  Counterintuitively, we find that  in suspensions with large volume fraction of nanoparticles and low ionic strength, the number of deprotonated groups can be 100\% larger in an isolated system, compared to a system connected to a reservoir by a semi-permeable membrane  -- both systems being at {\it exactly} the same pH.
		
	\end{abstract}
	\maketitle
	
	In many physics, chemistry, and biology applications one needs to understand systems with active surface groups~\cite{avni2020critical,monica1,podgornik2018,nishio1996monte,avni2019charge,lunkad2021role,da2009polyelectrolyte,Luijten,lund2005enhanced,Nap2014,Lundl,avni2019charge,majee2019,netz2002,da2009polyelectrolyte,2011xxiii,behrens2000charging, Ivan11}.  Depending on pH and salt concentration inside the solution, a surface group can be either protonated or deprotonated~\cite{Lund2013,lund2005charge,podgornik2018,landsgesell2019simulations,perez2021thermodynamic,Burak}.   The state in which a group finds itself depends on its intrinsic acid dissociation constant, the local concentration of hydronium ions, the electrostatic potential in the vicinity of the group, the steric repulsion between ions, etc.  For some simple surfaces  -- such as metal oxide or colloidal particles with regular arrangement of surface groups, it is possible to develop a theoretical approach which can fairly accurately predict the resulting surface charge and its dependence on salt and pH inside the solution~\cite{hiemstra1987proton,hiemstra1991physical,bakhshandeh2020interaction,bkh2022}.  Unfortunately no purely theoretical approach is possible for more complex systems, such as polyelectrolytes or proteins.  Furthermore, even in the case of simple surfaces in solutions containing multivalent ions  one finds that theoretical approaches begin to fail~\cite{Le02,levin1996criticality}.  For such complex systems one is forced to rely on computer simulations~\cite{labbez2006new,Labbezl,landsgesell2019simulations,Levincomment}. The constant pH (cpH) simulations method is the most popular tool to obtain titration curves of such systems~\cite{smith1994reaction,johnson1994reactive}.  There is, however, a fundamental difficulty in using such approach indiscriminately.  In cpH simulations,  proteins, colloidal particles, or polyelectrolyte,  are confined inside a simulation box, while ions can freely exchanged with the reservoir of acid and salt~\cite{monica2,Ong,wenncell}.  The cpH simulation method is, therefore, intrinsically semi-grand canonical~\cite{Levincomment}.  When performing a cpH simulation,  a proton is implicitly brought into the system from an external reservoir at a fixed pH.   To keep the charge neutrality inside the simulation cell one of the cations or protons inside the bulk of the cell is arbitrarily deleted.  However, such an arbitrary deletion does not respect the detailed balance and leads to incorrect results, unless the system contains a lot of salt and is very dilute in polyelectrolyte~\cite{labbez2006new,Levincomment}.  Fortunately, it is easy to correct the standard cpH algorithm by combining a protonation move with a simultaneous grand canonical insertion of an anion, and a deprotonation move with a simultaneous grand canonical deletion of an anion~\cite{labbez2006new,Labbezl,landsgesell2019simulations,Levincomment,lan,curk1}.  This restores the detailed balance of the cpH algorithm making it internally consistent.  This corrected version of the cpH algorithm will be used in the present Communication.
	
	In a real physical system, to confine  polyelectrolyte or colloidal particles within some region of space requires a semi-permeable membrane, which would separate the system from the reservoir containing acid and salt.  Since the counterions are at larger concentration inside the system than they are in the reservoir, they will tend to diffuse across the membrane~\cite{C6CS00632A}.  The net ionic flux will end when a sufficiently large electrostatic potential difference is established between the reservoir and the system, forcing the reversal of the flow.  This is known as the Donnan potential~\cite{donnan1911theorie,barr2012grand}.   In fact one does not need to have a membrane to establish a finite Donnan potential.  For example, a colloidal column in a  gravitational field~\cite{van2003defying,philipse2004remarks,warren2004electrifying} will spontaneously establish a finite Donnan potential along the vertical direction.  This happens because, in general,  colloidal particles have a finite buoyant mass~\cite{eckert2022sedimentation,sullivan2002control}, which forces  them to be inhomgeneously distributed inside the suspension.  On the other hand, the ions are not affected by the gravitational field and are free to diffuse throughout all space, including the top portion of suspension, where there are no colloidal particles.  This results in  Donnan potential along the vertical direction of  colloidal suspension.  In this case the gravitational field plays the role similar to that of a membrane, separating the region in which colloidal particles can be found.  The important point that is often lost when performing constant pH simulations is that the systems (simulation cell) is always at a different electrostatic potential from the reservoir.  On the other hand the electrochemical potentials, and consequently the pH,  inside the system and the reservoir are the same.

	Recall that pH of a solution is defined in terms of activity of hydronium ions, pH=$-\log_{10}\left[a_+/c^\ominus\right]$, where $c^\ominus=1$M is the standard concentration.  Inside the reservoir, the activity of hydronium ions is $a_+ = c_{\text{H}^+} \exp(\beta \mu_{ex}^r)$  where  $c_{\text{H}^+}$ is the concentration of hydroniums inside the reservoir and $\mu_{ex}$ is their excess chemical potential.  
	On the other hand inside the system $a_+ = \rho_{\text{H}^+} \exp[\beta \mu_{ex}^s + q \beta \varphi_D ]$, where $q$ is the proton charge, $\rho_{\text{H}^+}$ is the concentration of hydronium ions inside the system, and  $\varphi_D$ is the Donnan potential difference between the reservoir and the system.  Note that the concentration of hydroniums inside the system is different from their concentration in the reservoir.  Furthermore, while inside the reservoir  $\mu_{ex}^r$ is due only to the interaction between the ions, inside the system, $ \mu_{ex}^s$, also includes contributions coming from electrostatic and steric interactions of hydronium ion  with  the colloidal particles.
	Nevertheless, equivalence between electrochemical potentials inside the system and the reservoir requires that  pH of both be the same.  
	We see the important role that the Donnan potential plays for determining pH inside a system that is in contact with a reservoir.  If one will use hydrogen and calomel (reference) electrodes to measure the pH inside a system connected to a reservoir by a semi-permeable membrane -- with calomel placed in the reservoir -- the Donnan potential will be an integral part of the measurement.  Suppose now that the system is equilibrated and then disconnected from the reservoir  -- so that the number of ions of each type remains fixed and no longer fluctuates.  Clearly based on the ensemble equivalence of statistical mechanics, the number of protonated groups in such a {\it canonical} system will remain exactly the same as in the original semi-grand canonical system.  However, the pH of such a canonical system will no longer be the same as that of the semi-grand canonical system connected to the reservoir.  The reason for this is that inside 
	canonical system the potential drop between system and reservoir will no longer exist and activity of hydronium ions will be  $a_+ = \rho_{\text{H}^+} \exp[\beta \mu_{ex}^s]$.  Substituting this into the definition of pH we find the relation between the canonical and grand canonical pH:
	\begin{equation}
		\text{pH}_c = \text{pH}_{gc} +\frac{\beta\varphi_D}{\ln(10)}.
		\label{ph}
	\end{equation}
	
	Eq.  (\ref{ph}) opens a way for us to use the grand-canonical Monte Carlo (GCMC) simulation to simultaneously calculate the titration curves for both canonical and grand-canonical systems.  However to do this the usual GCMC algorithm, which relies on cation-anion pair insertion into the simulation cell, in order to preserve the charge neutrality, must be modified to allow for the charge fluctuation inside the cell, in order to calculate the Donnan potential.  Note that the Donnan potential is not accessible in the normal GCMC since it cancels out for pair moves involving oppositely charged particles.    In this Communication we will present a reactive GCMC-Donnan (rGCMCD) simulation method  which will allow us to calculate the number of protonated groups and the Donnan potential in a single simulation.  Combining this with Eq. (\ref{ph}) we are able to obtain the number of protonated group for a given pH in both canonical and grand-canonical systems.    We should stress that at this time there is no alternative method which permits us to calculate titration curves for isolated (canonical) systems.  In principle, one could use canonical reactive MC simulations and  then attempt to use Widom insertion method to obtain the excess chemical potential of hydronium ions inside the simulation cell.  The problem, however, is that for moderate and large pH there might not be any  hydronium ions inside the simulation cell, thus preventing us from obtaining an accurate canonical  pH.

	The difficulty in performing simulations of Coulomb systems stems from the long range electrostatic interaction between the particles that can not be cutoff at the cell boundary.  Instead, one is forced to periodically replicate the whole simulation cell, so that a given ion interacts not only with the ions inside the cell, but also with their infinite replicas.  For Donnan simulations presented in this Communication,  there is an additional difficulty because the simulation cell is no longer charge neutral.  The total charge inside the cell,  $Q_t=\sum_i q_i$, where the sum is over all the ions and the charged groups, is not necessarily zero at a given instant of simulation.  Clearly, such imbalance of charge in an infinitely replicated system leads to the divergence of the electrostatic energy.  To avoid this, we introduce a uniform neutralizing background of charge density $\rho_b=-Q_t/V$, where $V$ is the volume of the simulation cell.  The charge density inside the system can then be written as:
	
	\begin{eqnarray}\label{eq8a1}
		\rho_q({\pmb r}) = \sum_i q_i \delta({\pmb{r}}-{\pmb{r}}_i) -\frac{Q_t}{V}.
	\end{eqnarray}
	
	Using the usual approach, the electrostatic potential can be split into long and short range contribution~\cite{smith1981electrostatic,frenkel2001understanding,allen2017computer}.  
	The long range can be efficiently summed in the Fourier  space, while the short range is summed in the real space~\cite{Zhonghan,dos2016simulations,Shasha}, resulting in the electrostatic potential at position ${\pmb r}$ inside the simulation cell
	\begin{eqnarray}\label{eqq}
		\varphi({\pmb r})&=&\sum_{{\pmb k}={\pmb 0}}^{\infty}\sum_{j=1}^{N}\frac{4\pi \text{q}^j}{\epsilon_w V |{\pmb k}|^2}\exp{[-\frac{|{\pmb k}|^2}{4\kappa_e^2}+i{\pmb k}\cdot({\pmb r}-{\pmb r}^j)]} + \nonumber \\
		&&\sum_{j=1}^{N}\sum_{{\pmb n}}\text{q}^j\frac{\text{erfc}(\kappa_e|{\pmb r}-{\pmb r}^j-L{\pmb n}|)}{\epsilon_w |{\pmb r}-{\pmb r}^j-L{\pmb n}|} \nonumber \\
		&& + \frac{1}{V}\sum_{{\pmb k}={\pmb 0}}^{\infty} \tilde \varphi_b({\pmb k}) \exp[{i{\pmb k}\cdot{\pmb r}]}\ ,
	\end{eqnarray}
	where ${\pmb n}=(n_1,n_2,n_3)$ are integers,   ${\pmb k}=(\frac{2\pi}{L}n_1,\frac{2\pi}{L}n_2,\frac{2\pi}{L}n_3)$ are the reciprocal lattice vectors and $\kappa_e$ is  an arbitrary damping parameter.  The last term is the electrostatic potential due to the uniform 
	background,  the Fourier transform  of which is 
	\begin{equation}\label{ex}
		\tilde \varphi_b({\pmb k})= -\frac{4 \pi Q_t}{\epsilon_w V } \frac{\int_V   \mathrm{e}^{-i \pmb{k}.\pmb{r} }d^3r}{k^2}
	\end{equation}
	
	Note that the electrostatic potential, Eq. (\ref{eqq}), is invariant to the specific value of the damping parameter $\kappa_e$.  It is usual to choose  $\kappa_e$ sufficiently large, so that simple periodic boundary conditions can be used for the short range  part of the electrostatic potential.  
	
	The limit ${\pmb k} \rightarrow 0 $ in the Fourier sum of Eq. (\ref{eqq}) is singular and must be performed with great care.  The limiting procedure is shown in Appendix A, where 
	we obtain:
	\begin{eqnarray}\label{phib}
		\varphi({\pmb r})=\sum_{{\pmb k \neq 0}}^{\infty}\sum_{j=1}^{N}\frac{4\pi \text{q}^j}{\epsilon_w V |{\pmb k}|^2}\exp{[-\frac{|{\pmb k}|^2}{4\kappa_e^2}+i{\pmb k}\cdot({\pmb r}-{\pmb r}^j)]}  \nonumber \\
		+\sum_{j=1}^{N}\sum_{{\pmb n}}\text{q}^j\frac{\text{erfc}(\kappa_e|{\pmb r}-{\pmb r}^j -L{\pmb n} |)}{\epsilon_w |{\pmb r}-{\pmb r}^j-L{\pmb n}|} \\ \nonumber \  -\frac{ \pi Q_t}{\epsilon_w V\kappa_e^2}  +\frac{4 \pi}{3 \epsilon_w V }{\pmb{r}}\cdot{\pmb{M}} + \phi_B , \ 
	\end{eqnarray}
	where ${\pmb{M}} = \sum_i q_i {\pmb{r}}_i$ is the electric dipole moment inside the cell and $\phi_B$  is the modified Bethe potential of the cell with a neutralizing background:
	\begin{equation}\label{eqRM1}
		\phi_B = -\frac{2  \pi }{3 \epsilon_w V}\sum_i q_i {\pmb{r}_i}^2 + \frac{\pi Q_t}{6 \epsilon_w L}
	\end{equation}	 
	
\begin{figure}[H]
		\centering
		\includegraphics[width=0.4\linewidth]{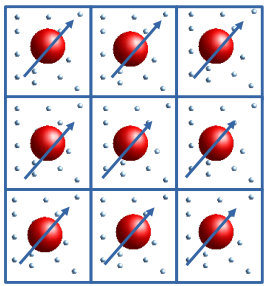}
		\caption{The replicas of the simulation cell. The large sphere is a colloidal particle and small spheres are ions, the arrows show the instantaneous electric dipole moment ${\pmb{M}}$, which in general exists inside the cell. Furthermore, the trace of the  instantaneous  second moment charge  distribution tensor of a locally strongly inhomogeneous system of colloidal particles and ions, will not be zero. A macroscopic spherical crystal produced by a periodic replication of the simulation cell will then have both surface charge density ${\pmb M} \cdot {\pmb n}/V$ and a dipolar layer, which results in a potential drop across the crystal--reservoir interface, given by the modified Bethe potential, $\phi_B$.}
		\label{fig:f1}
	\end{figure}

	We recognize the ${\pmb M}$ dependent term in Eq. (\ref{phib}) as arising from the uniform polarization of the macroscopic crystal composed from spherically replicated simulation cells.  From continuum electrostatics such uniform polarization is equivalent to the  surface charge density  ${\pmb M} \cdot {\pmb n}/V$,  where ${\pmb n}$ is the unit normal to the boundary of the macroscopic spherical crystal, resulting in an electrostatic potential $\frac{4 \pi}{3 \epsilon_w V }{\pmb{r}}\cdot{\pmb{M}}$ in the interior of the crystal, see Fig.~\ref{fig:f1}. 	
	The nature of  $\phi_B$ is more subtle.  In general, a simulation cell containing colloidal particles or polyelectrolyte,   will have a non-zero trace of the second moment charge density tensor.  Spherical replication of such cells will lead to a crystal with a dipole surface layer~\cite{euwema1975absolute}, across which there will be a potential drop precisely given by Eq. (\ref{eqRM1}).  In Appendix B, we also provide an alternative real space derivation of Eq. (\ref{eqRM1}).
	
	{\color{black}
		Since ions can diffuse across the membrane while colloidal particles are confined within the system, a Donnan potential exists between the reservoir and the system. While the Bethe potential can be thought of as the average of the local electrostatic potential inside the periodically replicated crystal, the Donnan potential is inherently an interfacial effect that arises from ionic diffusion between the system and the reservoir. The total electrostatic potential drop between the system and the reservoir is shown in Fig.~\ref{fn}.
		\begin{figure}[H]
			\centering
			\includegraphics[width=0.7\linewidth]{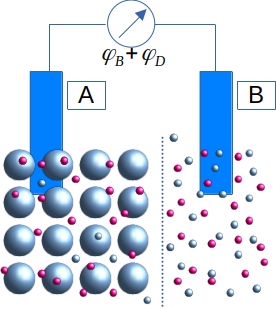}
			\caption{Periodically replicated system separated from the reservoir by a semi-permeable membrane: the electrostatic potential difference between the two is $\phi_B +\varphi_D$.}
			\label{fn}
		\end{figure}
		
	}
	The electrostatic energy can be calculate using 
	\begin{eqnarray}\label{eqE}
		&&E=\frac{1}{2} \int \rho_q({\pmb r})  \varphi({\pmb r}) d^3 {\pmb r} \nonumber \\
		&&=\frac{1}{2} \sum_i q_i \lim_{{\pmb r} \rightarrow {\pmb r}_i}\left[\varphi({\pmb r}) -\frac{q_i}{\epsilon_w |{\pmb r}-{\pmb r}_i|}\right]-\frac{Q_t}{2 V} \lim_{{\pmb k} \rightarrow {\pmb 0}} \tilde \varphi({\pmb k}).
	\end{eqnarray} 
	The subtraction inside the square brackets eliminates the self energy contribution to the total electrostatic energy of all ions and charged groups. The last term is due to the background charge.  Noting that the electrostatic energy is invariant with the damping parameter $\kappa_e$, the limit in the last term can be calculated using $\kappa_e \rightarrow \infty$, which kills off the real space sum in Eq. (\ref{phib}).  Taking the limit $\lim_{{\pmb k} \rightarrow {\pmb 0}}$,  the last term of Eq. (\ref{eqE}) then evaluates to  $Q_t \phi_B/2$  and the total electrostatic energy of a systems  with ions+sites of total charge $Q_t$ and a neutralizing background can be written as:
	\begin{eqnarray}\label{e6}
		&&E = \frac{1}{2}\sideset{}{'}\sum_{i j} \sum_{\pmb n}\frac{q_iq_j \text{erfc}(\kappa_e |{\pmb r}_i-{\pmb r}_j-L{\pmb n} |) }{\epsilon_w |{\pmb r}_i-{\pmb r}_j-L{\pmb n} |} \nonumber  \\
		&&+  \sum_{k\ne0} \frac{2 \pi\text{exp}(-{\pmb k}^2/4\kappa_e)}{\epsilon_wV{\pmb k}^2}(A({\pmb k})^2 + B({\pmb k})^2)  \nonumber \\
		&&-\sum_i\frac{q_i^2\kappa_e}{\epsilon_w \sqrt{\pi}} -\frac{\pi  Q_t^2}{2\epsilon_w V\kappa_e^2}+\frac{2 \pi}{3\epsilon_w V}  {\pmb{M}} ^2  ,
	\end{eqnarray}
	where
	\begin{eqnarray}\label{e61}
		A({\pmb k}) = \sum_i q_i\cos\left({\pmb k}\cdot {\pmb r}_i\right),\\ \nonumber
		B({\pmb k}) = \sum_i q_i\sin\left({\pmb k}\cdot {\pmb r}_i\right),\\ \nonumber
	\end{eqnarray}
	where the prime on the sum indicates that $i = j$ is excluded from the summation when  ${\bf n}=0$.
	Note that the Bethe potential canceled out and  does not contribute to the total electrostatic energy.
	Furthermore, since the system is implicitly connected to a reservoir, it is not possible to use the tinfoil boundary condition~\cite{de1981computer}  --  which would 
	eliminates ${\pmb k}={\pmb 0}$ term from the Ewald summation, isolating the system from the reservoir.  
	
	The simulation involves individual insertion/deletion moves, as well as protonation and deprotonation moves.
	For an ion of charge $q_i$ to enter the simulation cell, it must cross the macroscopic boundary of the crystal.  This will result in electrostatic energy change $q_i (\varphi_D +\phi_B)$.  The grand canonical acceptance probabilities for insertion/deletion moves can then be written as  $  \min\left(1,\phi_{add/rem}\right) $, where $\phi_{add/rem}$ is
	\begin{equation}\label{eqd1}
		\begin{split}
			\phi_{add} =  \frac{ c^\ominus  V 10^{-\text{pX}_i} }{N_i +1}\mathrm{e}^{ - \beta\left( \Delta E_{ele}  + q_i[\varphi_D +\phi_B]   \right)  },\\
			\phi_{rem} =  \frac{  N_i 10^{ \text{pX}_i}}{V  c^\ominus}\mathrm{e}^{- \beta\left( \Delta E_{ele}  - q_i [\varphi_D +\phi_B]   \right) }, 
		\end{split}
	\end{equation}
	where $\Delta E_{ele}$ is the change in electrostatic energy upon insertion/deletion of ion of type $i$,  
	$\text{pX}_i\equiv -\log_{10} X_i/c^\ominus$, where $X_i$ is the activity of ion $i$ inside the reservoir. To relate $\text{pX}_i$ to the concentration of salt, one can perform a separate GCMC simulation for the reservoir.  In practice, however, we find that for monovalent ions the mean spherical approximation (MSA) combined with Carnahan-Starling (CS) expression for the excluded volume~\cite{ho1988mean,carnahan1969equation} provides an excellent approximation for the relation between $\text{pX}_i$, $\text{pH}$ and the concentrations of salt and acid in the reservoir.
	
	The protonation/deprotonation moves involve a reaction 
	\begin{eqnarray}\label{eqRM}
		\text{H}\text{A}\rightleftharpoons 	\text{H}_3\text{O}^+ + \text{A}^{-}, \\ \nonumber  
		K_a = \frac{a_{\text{A}^-}a_{\text{H}^+}}{a_{\text{HA}}},
	\end{eqnarray}
	where $\text{A}^{-}$ is the deprotonated site.  We recognize the acid dissociation constant to be the inverse of the internal partition function of HA molecule.
	The free energy change due removal of hydronium from the  reservoir and its reaction with an {\it isolated} A$^-$ group is then
	\begin{equation}\label{eqRM}
		\beta \Delta F_p =   \ln\left( \frac{K_{a}}{c^\ominus} \right)   - \mu_{\text{H}^+},
	\end{equation}
	where $  \mu_{\text{H}^+} = \ln(c_{\text{H}^+}/c^\ominus) + \beta\mu_{ex}$ is the total chemical
	potential of a hydronium ion inside the reservoir. For
	the deprotonation reaction the free energy change is reversed, so that $\Delta F_d  = -\Delta F_p$.  
	The acceptance probabilities for protonation and deprotonation moves are then $\min\left(1,\phi_{p/d}\right)$ where
	\begin{equation}\label{eqRM_1}
		\begin{split}
			\phi_p =  \mathrm{e}^{-\beta \left(\Delta E_{ele}  +  \Delta F_p    +q[\varphi_D +\phi_B]   \right)} =\\10^{ \text{pK}_a-\text{pH}_a}\mathrm{e}^{-\beta \left(\Delta E_{ele}      +q[\varphi_D +\phi_B]    \right)} ,\\
			\phi_d = \mathrm{e}^{-\beta \left(\Delta E_{ele} +\Delta F_d  -q[\varphi_D +\phi_B]      \right)}= \\10^{\text{pH} -\text{pK}_a} \mathrm{e}^{-\beta \left(\Delta E_{ele}   -q[\varphi_D +\phi_B]     \right)}.
		\end{split}
	\end{equation}
	We have defined the intrinsic $\text{pK}_a$ of a titration site as $\text{pK}_a=-\log_{10}[\text{K}_a/c^\ominus]$.
	During the protonation/deprotonation moves, a titration site is randomly chosen and its state is changed with acceptance probability give by Eq. (\ref{eqRM_1}). 
	
	The simulation starts with an initial guess for the value of  $\varphi_D$.  After each MC step, the Donnan potential is automatically updated, so as to drive the system toward a charge neutral state: $\varphi_D^{new} := \varphi_D^{old} + \alpha Q_t$,
	where  $\alpha>0$ is an arbitrary small parameter that controls convergence.  The simulation stops when $\langle Q_t\rangle \approx 0$, to the desired accuracy.  {\color{black} In practice the system very quickly converges to a state in which $Q_t$ has small oscillations about zero.  To make sure that the system is fully equilibrated, we monitor the energy.  When  $\langle E \rangle$ becomes constant, we know that the system has reached equilibrium.  Usually we use $5\times 10^6$ particle moves for equilibration.  After this we collect samples which include colloidal charge, Donnan potential, ion distribution, etc.  The samples are collected with intervals of approximately $3000$ particle moves,  to make sure that they are fully uncorrelated.  We use $20000$ such sample to calculate the averages.} 
	

		
		As an example of the algorithm developed in the present Communication, we will use it to study a colloidal suspension of finite volume fraction.  The colloidal particles have radius $60$~\AA, and contain $Z=600$ active surface groups of intrinsic pK$_{a}$=5.4 uniformly distributed over the surface.  All ions have radius $2$~\AA, and the solvent is modeled as a dielectric continuum of Bjerrum length, $\lambda_B=q^2/\epsilon k_B T=7.2$ \AA. 
		The reservoir contain acid of an arbitrary pH and 1:1 salt of concentration 1 mM.  Electrostatic energy is calculated using Eq. (\ref{e6}), with the damping parameter set to $\kappa_e=5/L$, where $L$ is the size of the cubic simulation box.  Such value of  $\kappa_e$ is sufficiently large that simple periodic boundary conditions can be used to evaluate the short range (sum over erfc functions) contribution to the electrostatic energy in Eq. (\ref{e6}).
		
		For simplicity in the present Communication we will use only one colloidal particle inside a cubic simulation box of $L=200$~\AA.  There is, however, no conceptual difficulty for using arbitrary number of colloidal particles inside a simulation box, except for the CPU time constraint. Finally, to test the new rGCMCD algorithm we perform a ``standard" cpH in which a protonation move is  combined with a grand canonical  insertion of anion and a deprotonation move with a grand canonical  deletion of an anion~\cite{labbez2006new,reactive,Levincomment} .   The acceptance probabilities for these pair moves are respectively:
		\begin{eqnarray}\label{eq8}
			&&\text{P}_p =\min\left[1,\frac{c^\ominus V 10^{\text{pK}_a-\text{pH} -\text{pCl}}}{ N_{\ch{Cl}}+1 }e^{-\beta \Delta E }\right]  \nonumber \\
			&&\text{P}_d=\min\left[1,\frac{ N_{\ch{Cl}} 10^{\text{pH} -  \text{pK}_a +\text{pCl}}}{c^\ominus V } e^{-\beta \Delta E}\right]. 
		\end{eqnarray}
		Since both Bethe and Donnan potential cancel in the pair moves, cpH simulation does not permit us to calculate the canonical titration curve.  On the other hand, the grand-canonical curve obtained using the ``standard" cpH must agree precisely with the one obtained using our new rGCMCD simulation method.  This is exactly what is found, validating the 
		rGCMCD method.
		
\begin{figure}[H]
			\centering
			\includegraphics[width=0.6\linewidth]{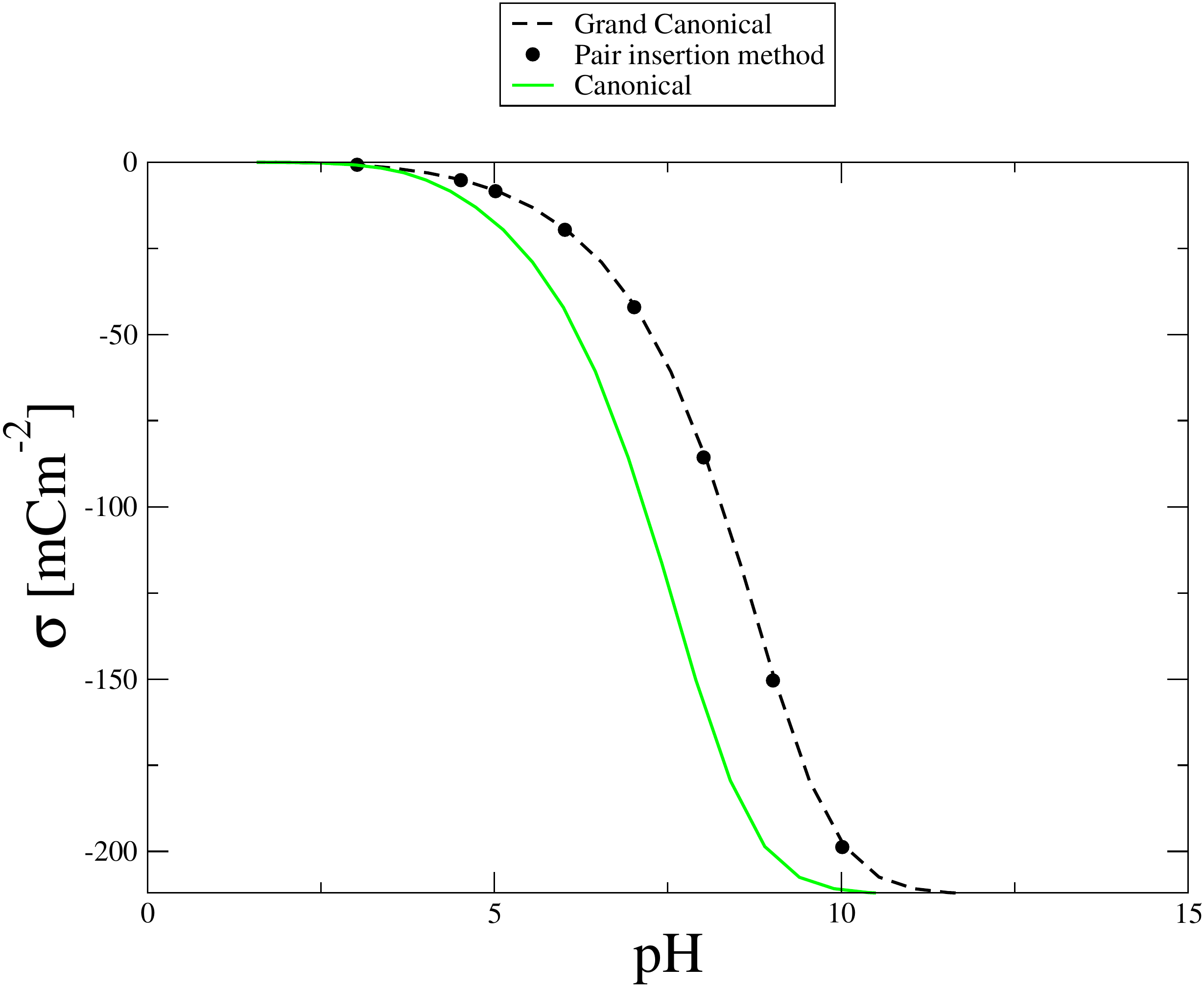}
			\caption{Titration curves for an isolated (canonical, solid curve) and semi-grand canonical system (dashed curve) in contact with a salt and acid reservoir, calculated using rGCMCD simulations. In  semi-grand canonical case the reservoir contains 1mM of salt,  for canonical case the concentration in reservoir is fine tuned,  so that there is 1mM of salt inside the system for all pH.  This, however, has an imperceptible effect on canonical titration curve for low salt concentrations.  Points are the results of the ``standard"  cpH  with pair insertions, which agree perfectly with the new rGCMCD method.   Colloidal suspension has volume fraction 11.3\%.  The saturated colloidal charge is -212mCm$^{-2}$. The pK$_a$ of acidic groups is 5.4.}
			\label{ffig1}
		\end{figure}

\begin{figure}[H]
			\centering
			\includegraphics[width=0.7\linewidth]{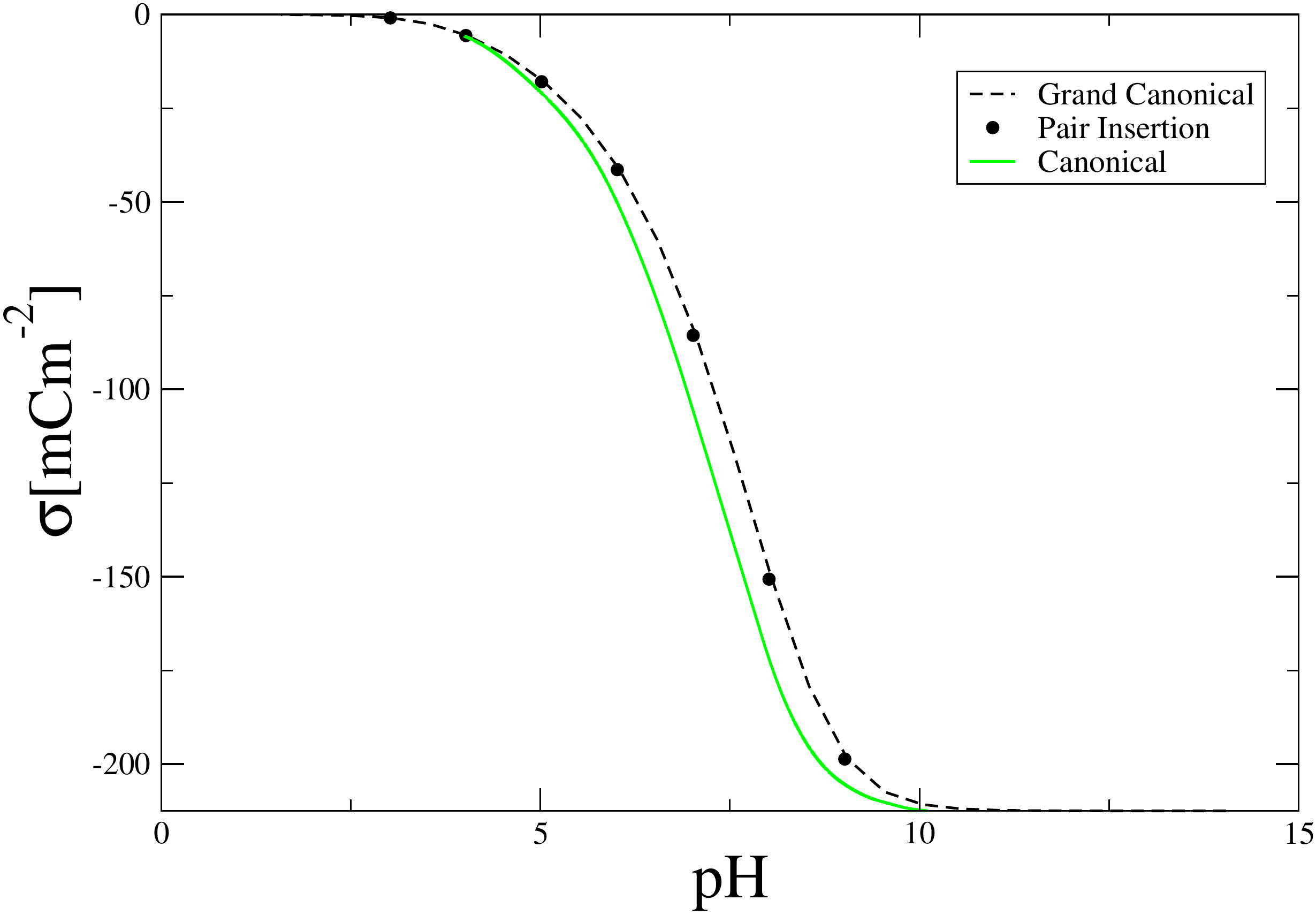}
			\caption{Titration curves for an isolated (canonical, solid curve) and semi-grand canonical system (dashed curve) in contact with a salt and acid reservoir, calculated using rGCMCD simulations. In  semi-grand canonical case the reservoir contains 10mM of salt,  for canonical case the concentration in reservoir is fine tuned,  so that there is 10mM of salt inside the system for all pH. Points are the results of cpH pair insertion simulation. Colloidal suspension has volume fraction 11.3\%.  The saturated colloidal surface charge is -212mCm$^{-2}$. The pK$_a$ of acidic groups is 5.4.}
			\label{fig2}
		\end{figure}

		In Fig.~\ref{ffig1} we present the titration curves obtained using our rGCMCD simulations for both canonical (isolated) and  semi-grand canonical suspensions.  
		As can be seen from the figure,  there is a very significant difference between titration curves of an isolated colloidal suspension and of suspension connected to a salt and acid reservoir. For example, at pH=7.5,  28\% of surface groups of semi-grand canonical system are deprotonated, while for an isolated canonical system this jumps to 57\%.  When the concentration of salt increases, the difference between canonical and semi-grand canonical systems diminishes, see Fig. \ref{fig2}. This justifies the use of cpH methods to study biologically relevant systems that contain physiological concentrations of salt of $c_s \approx 150$ mM, and which are usually isolated from any external reservoir. 
		
				We have presented a new simulation method that enables us to simultaneously obtain titration curves 
				both for isolated systems (canonical ensemble) and systems connected to an external reservoir by a semi-permeable membrane.  Counterintuitively, we find that at low ionic strength, the number of deprotonated groups can be 100\% larger in an isolated system, compared to a system connected to a reservoir by a semi-permeable membrane  ---  both systems at exactly the same pH!  The difference can be even greater for  more concentrated (large colloidal volume fraction) suspensions in deionized solutions.  As the concentration of salt increases, the difference between the ensembles becomes less important, justifying the use of cpH methods for studying biologically relevant systems that are usually closed and contain large concentrations of salt.  It is important to stress that at the moment there is no alternative method available to study canonical systems at moderate and high pH.  The usual canonical reactive simulations fail under such conditions, since the simulation cell  contains none or very few hydronium ions, preventing one from accurately calculating the pH inside the system.
				
				{\color{black} In our definition of pH for heterogeneous systems, we have followed the Gibbs-Guggenheim principle~\cite{GuggenheimII}, which prohibits separation of the electrochemical potential into chemical and electrostatic potential contributions for definition of measurable quantities. On the other hand, in electrochemistry literature Gibbs-Guggenheim principle is often disregarded and pH is defined in terms of only the chemical part of the total electrochemical potential.  
					From the practical point of view, pH is determined by measuring the electromotive force (EMF) between a glass or hydrogen electrode and a saturated calomel (reference) electrode. In our definition of pH, the reference electrode is always kept inside the reservoir, while the hydrogen electrode is moved between the reservoir and the system.  
					With this setup, we will obtain the same EMF independent of the position of the hydrogen
					electrode, indicating the same pH inside the system and in the reservoir~\cite{Teorell}.  
					On the other hand, if both electrodes are moved together -- changing the reference value of the electrostatic potential -- we will obtain different pH values inside the system and in the reservoir. With such definition, system's pH will be different from that  of the reservoir, but will be the same as of an isolated canonical system.  Whichever definition of pH is chosen, one needs to know the Donnan potential in order to calculate the titration isotherms of isolated (canonical) systems using semi-grand canonical simulations.  The ``standard" cpH simulation methods do not allow us to calculate titration isotherms for isolated systems.  One must either use canonical reactive MC -- with Widom particle insertion, which is very inaccurate even for intermediate to high values -- or the approach presented in the present Communication, which is accurate for any pH}.  
				
				\section{Acknowledgments}
				
				This work was partially supported by the CNPq, the CAPES, and the National Institute of Science and Technology Complex Fluids INCT-FCx.  
				
				\section{Conflict of interest}

				\appendix
				
				The authors have no conflicts to disclose.

				\section{Details for the limiting procedure for singular term of electrostatic potential}
				
				The electrostatic potential inside the simulation cell is: 
				\begin{eqnarray}\label{eqq}
					\phi({\pmb r})&=&\sum_{{\pmb k}={\pmb 0}}^{\infty}\sum_{j=1}^{N}\frac{4\pi \text{q}_j}{\epsilon_w V |{\pmb k}|^2}\exp{[-\frac{|{\pmb k}|^2}{4\kappa_e^2}+i{\pmb k}\cdot({\pmb r}-{\pmb r}_j)]} + \nonumber \\
					&&\sum_{j=1}^{N}\sum_{{\pmb n}}\text{q}_j\frac{\text{erfc}(\kappa_e|{\pmb r}-{\pmb r}_j-L{\pmb n}|)}{\epsilon_w |{\pmb r}-{\pmb r}_j-L{\pmb n}|} \nonumber \\
					&&+ \frac{1}{V}\sum_{{\pmb k}={\pmb 0}}^{\infty} \tilde \phi_b({\pmb k}) \exp[{i{\pmb k}\cdot{\pmb r}]}\ ,
				\end{eqnarray}
				where  $V=L^3$ is the volume of the cubic simulation cell,  $Q_t=\sum_i \text{q}_i$ and
				\begin{equation}\label{e1}
					\tilde \phi_b({\pmb k})= -\frac{4 \pi Q_t}{\epsilon_w V } \frac{\int_V   \mathrm{e}^{-i \pmb{k}.\pmb{r} }d^3r}{k^2},
				\end{equation}
				is the Fourier transform of the background potential. 
				Note that the integral in Eq. (\ref{e1}) vanishes for all ${\pmb k}$, except for ${\pmb k}={\pmb 0}$, so that we can define the singular part of the background  potential $ \phi_{b,s}$ as $\tilde \phi_b({\pmb k})\equiv \tilde \phi_{b,s} \delta_{{\pmb k}{\pmb 0}}$, where we have defined the Kronecker delta for the zero mode, $\delta_{{\pmb k}{\pmb 0}}$.
				As  ${\pmb k} \rightarrow 0$,  we can expand the singular part around  ${\pmb k}={\pmb 0}$,
				\begin{equation}\label{e2}
					\tilde \phi_{b,s}= -\frac{4 \pi Q_t }{\epsilon_w V } \frac{\int_V  \left(1-i \pmb{k} \cdot \pmb{r} - \frac{1}{2}(\pmb{k} \cdot \pmb{r})^2 +...\right) d^3r}{k^2}.
				\end{equation}
				The leading order terms in $k$ are:
				\begin{equation}\label{e3}
					\tilde \phi_{b,s}=  -\frac{4 \pi Q_t }{\epsilon_w k^2} +\frac{4 \pi Q_t}{\epsilon_w k^2 V}\int_V  i \pmb{k}.\pmb{r}   d^3r +  \frac{2 \pi Q_t}{\epsilon_w V  k^2}\int_V(\pmb{k}.\pmb{r})^2 d^3r, 
				\end{equation}
				with the higher order terms vanishing in the limit ${\pmb k} \rightarrow 0$.   We note that the limit ${\pmb k} \rightarrow 0$ corresponds to large distance behavior of the electrostatic potential.  To clearly see how the divergence scales with the size of the macroscopic crystal, $L_{cr}$,  produced by the periodic replication of simulation cell, see Fig. (\ref{fig:f1}), 
				we introduce a ``crystal" delta function:
				\begin{eqnarray}\label{dcr}
					&&\delta_{cr}({\pmb k})=\frac{1}{(2 \pi)^3}\prod_{i=1}^3\int_{-L_{cr}/2}^{L_{cr}/2} e^{i k_i p_i} dp_i= \nonumber\\
					&&\frac{1}{ \pi^3}\prod_{i=1}^3\frac{\sin(k_i  L_{cr}/2)}{k_i} \ .
				\end{eqnarray}
				In the limit $L_{cr} \rightarrow \infty$, $\delta_{cr}({\pmb k})$ becomes the Dirac delta function. The crystal $\delta_{cr}({\pmb k})$ regularizes the divergence and allows us to study the  ${\pmb k} \rightarrow 0$ limit~\cite{dos2016simulations}.
				In particular:
				\begin{eqnarray}\label{eqkk}
					\lim_{{\pmb k} \rightarrow 0} \frac{k_i k_j}{{\pmb k}^2}=\int \delta_{cr}({\pmb k})\frac{k_i k_j}{{\pmb k}^2} d^3 {\pmb k}=\frac{1}{3} \delta_{ij} \,
				\end{eqnarray}
				where $\delta_{ij}$ is the Kronecker delta.  The last equality follows from the direct use of the representation of $\delta_{cr}({\pmb k})$,  Eq. (\ref{dcr}), followed by $L_{cr} \rightarrow \infty$ limit~\cite{dos2016simulations}, or simply from the observation of symmetry.  
				Similarly,
				\begin{eqnarray}\label{eqk}
					\lim_{{\pmb k} \rightarrow 0} \frac{k_i }{{\pmb k}^2}=\int \delta_{cr}({\pmb k})\frac{k_i }{{\pmb k}^2} d^3 {\pmb k}=0 \,,
				\end{eqnarray}
				since $\delta_{cr}({\pmb k})$ is an even function of all $k_i$.
				Using Eq. (\ref{eqk}), we see that the second term of Eq. (\ref{e3}) vanishes and the last term becomes:
				\begin{eqnarray}\label{eqlas}
					\frac{2 \pi Q}{3 \epsilon_w V }\int_V( x^2+y^2+z^2) dx dy dz=\frac{ \pi Q L^2 }{6 \epsilon_w },
				\end{eqnarray}
				so that
				\begin{equation}\label{e4}
					\tilde \phi_{b,s}=-\frac{ 4\pi Q_t}{\epsilon_w k ^2}   + \frac{\pi Q_t L^2}{6 \epsilon_w }.
				\end{equation}

				We now expand the first term of Eq.~\ref{eqq} around  ${\pmb k}= 0$. The singular terms are: 
				\begin{eqnarray}\label{e5}
					\frac{4 \pi}{V \epsilon_w k^2}\sum_{j=1} \text{q}_j -\frac{\pi}{\epsilon_w V \kappa_e^2}\sum_{j=1}\text{q}_j+\nonumber \\
					\frac{4 \pi}{V \epsilon_w}\sum_{j=1} \text{q}_j\frac{i{\pmb k}\cdot({\pmb r}-{\pmb r}_j)}{|{\pmb k}|^2} - \nonumber \\ 
					\frac{2 \pi}{V \epsilon_w}
					\sum_{j=1} \text{q}^j\dfrac{[{\pmb k}\cdot({\pmb r}-{\pmb r}_j)]^2}{ |{\pmb k}|^2}  \ ,
				\end{eqnarray}
				which using Eqs (\ref{eqk}) and (\ref{eqkk}) simplify to: 
				\begin{eqnarray}\label{e6}
					\frac{4 \pi Q_t}{ \epsilon_w k^2 V} -\frac{\pi Q_t}{\epsilon_w V \kappa_e^2}
					- \frac{2 \pi}{3 V \epsilon_w}
					\sum_{j=1} \text{q}^j ({\pmb r}-{\pmb r}_j)^2 \ .
				\end{eqnarray}
				On the other hand, the expansion of the last term of Eq.~\ref{eqq} around  ${\pmb k}= 0$, with the  help of Eq. (\ref{e4}) results in:
				
				\begin{eqnarray}\label{e7}
					-\frac{ 4\pi Q_t}{\epsilon_w k ^2 V}   + \frac{\pi Q_t }{6 \epsilon_w L} + \frac{2 \pi Q_t r^2 }{3 V \epsilon_w} \ .
				\end{eqnarray}
				
				Combining everything, we see that the diverging $1/k^2$ terms cancel out and we obtain  the electrostatic potential 
				inside the simulation cell:
				\begin{eqnarray}\label{phib}
					&&\varphi({\pmb r})=\sum_{{\pmb k \neq 0}}^{\infty}\sum_{j=1}^{N }\frac{4\pi \text{q}_j}{\epsilon_w V |{\pmb k}|^2}\exp{[-\frac{|{\pmb k}|^2}{4\kappa_e^2}+i{\pmb k}\cdot({\pmb r}-{\pmb r}_j)]}  \nonumber \\
					&&+\sum_{j=1}^{N }\sum_{{\pmb n}}\text{q}_j\frac{\text{erfc}(\kappa_e|{\pmb r}-{\pmb r}_j -L{\pmb n} |)}{\epsilon_w |{\pmb r}-{\pmb r}_j-L{\pmb n}|} \\  &&- \frac{\pi Q_t}{\epsilon_w V\kappa_e^2} +\frac{4 \pi}{3 \epsilon_w V }{\pmb{r}}\cdot{\pmb{M}} -\frac{2 \pi}{3\epsilon_w V }\sum_i\text{q}_i {\pmb{r}}_i^2   +  \frac{\pi Q_t}{6 \epsilon_w L},  \nonumber 
				\end{eqnarray}
				
				\section{Real space derivation of the Bethe potential inside a cell with a neutralizing background.}
				
				The electrostatic potential, Eq. (\ref{phib}), is invariant with respect to $\kappa_e$.  In the limit $\kappa_e \rightarrow \infty$, the sum over ${\text{erfc}}$ vanishes, and the electrostatic potential inside the cell can be written  in terms of only the Fourier components.  
				It is clear then that 
				\begin{eqnarray}\label{int}
					\frac{1}{V} \int_V \varphi({\pmb r}) d^3 {\pmb r}=\phi_B .
				\end{eqnarray}
				Following Bethe, we can also calculate this as the mean potential over the whole spherical crystal of radius $V_{cr}=4 \pi R^3/3$,
				\begin{eqnarray}\label{bet}
					\phi_B=\frac{1}{\epsilon_w V_{cr}} \int_{V_{cr}} \int_{V_{cr}} \frac{\rho_q({\pmb r}')}{|{\pmb r}'-{\pmb r} |} d^3 {\pmb r'}d^3 {\pmb r}.
				\end{eqnarray}
				Interchanging the orders of integration, the integral over ${\pmb r}$ is the potential at position ${\pmb r'}$ inside a uniformly charged sphere of unit charge density, which can be easily obtained from the solution of Poisson equation:
				\begin{eqnarray}\label{bet2}
					\int_{V_{cr}} \frac{1}{|{\pmb r}'-{\pmb r} |} d^3 {\pmb r}=\frac{2 \pi}{3} \left( 3 R^2 -r'^2 \right).
				\end{eqnarray}
				The average potential inside the simulation cell is then:
				\begin{eqnarray}\label{bet3}
					\phi_B=\frac{2 \pi}{3V}\int_V  \left( 3 R^2 -r'^2 \right) \rho_q({\pmb r}') d^3 {\pmb r'}= \nonumber \\
					-\frac{2 \pi}{3V}\int_V  r'^2  \rho_q({\pmb r}') d^3 {\pmb r'},
				\end{eqnarray}
				where in the last equality we have used the total ions+sites+background charge neutrality inside the cell.  Finally, substituting $\rho_q({\pmb r}) = \sum_i q_i \delta({\pmb{r}}-{\pmb{r}}_i) -\frac{Q_t}{V}$ into the equation above, we obtain:
				\begin{eqnarray}\label{bet3}
					\phi_B=-\frac{2 \pi}{3\epsilon_w V }\sum_i\text{q}_i {\pmb{r}}_i^2   +  \frac{\pi Q_t}{6 \epsilon_w L}.
				\end{eqnarray}
				\bibliography{ref}

\begin{thebibliography}{58}%
\makeatletter
\providecommand \@ifxundefined [1]{%
 \@ifx{#1\undefined}
}%
\providecommand \@ifnum [1]{%
 \ifnum #1\expandafter \@firstoftwo
 \else \expandafter \@secondoftwo
 \fi
}%
\providecommand \@ifx [1]{%
 \ifx #1\expandafter \@firstoftwo
 \else \expandafter \@secondoftwo
 \fi
}%
\providecommand \natexlab [1]{#1}%
\providecommand \enquote  [1]{``#1''}%
\providecommand \bibnamefont  [1]{#1}%
\providecommand \bibfnamefont [1]{#1}%
\providecommand \citenamefont [1]{#1}%
\providecommand \href@noop [0]{\@secondoftwo}%
\providecommand \href [0]{\begingroup \@sanitize@url \@href}%
\providecommand \@href[1]{\@@startlink{#1}\@@href}%
\providecommand \@@href[1]{\endgroup#1\@@endlink}%
\providecommand \@sanitize@url [0]{\catcode `\\12\catcode `\$12\catcode
  `\&12\catcode `\#12\catcode `\^12\catcode `\_12\catcode `\%12\relax}%
\providecommand \@@startlink[1]{}%
\providecommand \@@endlink[0]{}%
\providecommand \url  [0]{\begingroup\@sanitize@url \@url }%
\providecommand \@url [1]{\endgroup\@href {#1}{\urlprefix }}%
\providecommand \urlprefix  [0]{URL }%
\providecommand \Eprint [0]{\href }%
\providecommand \doibase [0]{http://dx.doi.org/}%
\providecommand \selectlanguage [0]{\@gobble}%
\providecommand \bibinfo  [0]{\@secondoftwo}%
\providecommand \bibfield  [0]{\@secondoftwo}%
\providecommand \translation [1]{[#1]}%
\providecommand \BibitemOpen [0]{}%
\providecommand \bibitemStop [0]{}%
\providecommand \bibitemNoStop [0]{.\EOS\space}%
\providecommand \EOS [0]{\spacefactor3000\relax}%
\providecommand \BibitemShut  [1]{\csname bibitem#1\endcsname}%
\let\auto@bib@innerbib\@empty
\bibitem [{\citenamefont {Avni}, \citenamefont {Podgornik},\ and\ \citenamefont
  {Andelman}(2020)}]{avni2020critical}%
  \BibitemOpen
  \bibfield  {author} {\bibinfo {author} {\bibfnamefont {Y.}~\bibnamefont
  {Avni}}, \bibinfo {author} {\bibfnamefont {R.}~\bibnamefont {Podgornik}}, \
  and\ \bibinfo {author} {\bibfnamefont {D.}~\bibnamefont {Andelman}},\
  }\bibfield  {title} {\enquote {\bibinfo {title} {Critical behavior of
  charge-regulated macro-ions},}\ }\href@noop {} {\bibfield  {journal}
  {\bibinfo  {journal} {The Journal of Chemical Physics}\ }\textbf {\bibinfo
  {volume} {153}},\ \bibinfo {pages} {024901} (\bibinfo {year}
  {2020})}\BibitemShut {NoStop}%
\bibitem [{\citenamefont {Prusty}\ \emph {et~al.}(2020)\citenamefont {Prusty},
  \citenamefont {Nap}, \citenamefont {Szleifer},\ and\ \citenamefont {Olvera
  de~la Cruz}}]{monica1}%
  \BibitemOpen
  \bibfield  {author} {\bibinfo {author} {\bibfnamefont {D.}~\bibnamefont
  {Prusty}}, \bibinfo {author} {\bibfnamefont {R.~J.}\ \bibnamefont {Nap}},
  \bibinfo {author} {\bibfnamefont {I.}~\bibnamefont {Szleifer}}, \ and\
  \bibinfo {author} {\bibfnamefont {M.}~\bibnamefont {Olvera de~la Cruz}},\
  }\bibfield  {title} {\enquote {\bibinfo {title} {Charge regulation mechanism
  in end-tethered weak polyampholytes},}\ }\href@noop {} {\bibfield  {journal}
  {\bibinfo  {journal} {Soft Matter}\ }\textbf {\bibinfo {volume} {16}},\
  \bibinfo {pages} {8832--8847} (\bibinfo {year} {2020})}\BibitemShut {NoStop}%
\bibitem [{\citenamefont {Podgornik}(2018)}]{podgornik2018}%
  \BibitemOpen
  \bibfield  {author} {\bibinfo {author} {\bibfnamefont {R.}~\bibnamefont
  {Podgornik}},\ }\bibfield  {title} {\enquote {\bibinfo {title} {General
  theory of charge regulation and surface differential capacitance},}\
  }\href@noop {} {\bibfield  {journal} {\bibinfo  {journal} {J. Chem. Phys.}\
  }\textbf {\bibinfo {volume} {149}},\ \bibinfo {pages} {104701} (\bibinfo
  {year} {2018})}\BibitemShut {NoStop}%
\bibitem [{\citenamefont {Nishio}(1996)}]{nishio1996monte}%
  \BibitemOpen
  \bibfield  {author} {\bibinfo {author} {\bibfnamefont {T.}~\bibnamefont
  {Nishio}},\ }\bibfield  {title} {\enquote {\bibinfo {title} {Monte carlo
  studies on potentiometric titration of (carboxymethyl) cellulose},}\
  }\href@noop {} {\bibfield  {journal} {\bibinfo  {journal} {Biophysical
  chemistry}\ }\textbf {\bibinfo {volume} {57}},\ \bibinfo {pages} {261--267}
  (\bibinfo {year} {1996})}\BibitemShut {NoStop}%
\bibitem [{\citenamefont {Avni}, \citenamefont {Andelman},\ and\ \citenamefont
  {Podgornik}(2019)}]{avni2019charge}%
  \BibitemOpen
  \bibfield  {author} {\bibinfo {author} {\bibfnamefont {Y.}~\bibnamefont
  {Avni}}, \bibinfo {author} {\bibfnamefont {D.}~\bibnamefont {Andelman}}, \
  and\ \bibinfo {author} {\bibfnamefont {R.}~\bibnamefont {Podgornik}},\
  }\bibfield  {title} {\enquote {\bibinfo {title} {Charge regulation with fixed
  and mobile charged macromolecules},}\ }\href@noop {} {\bibfield  {journal}
  {\bibinfo  {journal} {Curr Opin Electrochem}\ }\textbf {\bibinfo {volume}
  {13}},\ \bibinfo {pages} {70--77} (\bibinfo {year} {2019})}\BibitemShut
  {NoStop}%
\bibitem [{\citenamefont {Lunkad}\ \emph {et~al.}(2021)\citenamefont {Lunkad},
  \citenamefont {Murmiliuk}, \citenamefont {To{\v{s}}ner}, \citenamefont
  {{\v{S}}t{\v{e}}p{\'a}nek},\ and\ \citenamefont
  {Ko{\v{s}}ovan}}]{lunkad2021role}%
  \BibitemOpen
  \bibfield  {author} {\bibinfo {author} {\bibfnamefont {R.}~\bibnamefont
  {Lunkad}}, \bibinfo {author} {\bibfnamefont {A.}~\bibnamefont {Murmiliuk}},
  \bibinfo {author} {\bibfnamefont {Z.}~\bibnamefont {To{\v{s}}ner}}, \bibinfo
  {author} {\bibfnamefont {M.}~\bibnamefont {{\v{S}}t{\v{e}}p{\'a}nek}}, \ and\
  \bibinfo {author} {\bibfnamefont {P.}~\bibnamefont {Ko{\v{s}}ovan}},\
  }\bibfield  {title} {\enquote {\bibinfo {title} {Role of p ka in charge
  regulation and conformation of various peptide sequences},}\ }\href@noop {}
  {\bibfield  {journal} {\bibinfo  {journal} {Polymers}\ }\textbf {\bibinfo
  {volume} {13}},\ \bibinfo {pages} {214} (\bibinfo {year} {2021})}\BibitemShut
  {NoStop}%
\bibitem [{\citenamefont {da~Silva}\ and\ \citenamefont
  {J{\"o}nsson}(2009)}]{da2009polyelectrolyte}%
  \BibitemOpen
  \bibfield  {author} {\bibinfo {author} {\bibfnamefont {F.~L.~B.}\
  \bibnamefont {da~Silva}}\ and\ \bibinfo {author} {\bibfnamefont
  {B.}~\bibnamefont {J{\"o}nsson}},\ }\bibfield  {title} {\enquote {\bibinfo
  {title} {Polyelectrolyte--protein complexation driven by charge
  regulation},}\ }\href@noop {} {\bibfield  {journal} {\bibinfo  {journal}
  {Soft Matter}\ }\textbf {\bibinfo {volume} {5}},\ \bibinfo {pages}
  {2862--2868} (\bibinfo {year} {2009})}\BibitemShut {NoStop}%
\bibitem [{\citenamefont {Curk}\ and\ \citenamefont {Luijten}(2021)}]{Luijten}%
  \BibitemOpen
  \bibfield  {author} {\bibinfo {author} {\bibfnamefont {T.}~\bibnamefont
  {Curk}}\ and\ \bibinfo {author} {\bibfnamefont {E.}~\bibnamefont {Luijten}},\
  }\bibfield  {title} {\enquote {\bibinfo {title} {Charge regulation effects in
  nanoparticle self-assembly},}\ }\href@noop {} {\bibfield  {journal} {\bibinfo
   {journal} {Phys. Rev. Lett.}\ }\textbf {\bibinfo {volume} {126}},\ \bibinfo
  {pages} {138003} (\bibinfo {year} {2021})}\BibitemShut {NoStop}%
\bibitem [{\citenamefont {Lund}, \citenamefont {{\AA}kesson},\ and\
  \citenamefont {J{\"o}nsson}(2005)}]{lund2005enhanced}%
  \BibitemOpen
  \bibfield  {author} {\bibinfo {author} {\bibfnamefont {M.}~\bibnamefont
  {Lund}}, \bibinfo {author} {\bibfnamefont {T.}~\bibnamefont {{\AA}kesson}}, \
  and\ \bibinfo {author} {\bibfnamefont {B.}~\bibnamefont {J{\"o}nsson}},\
  }\bibfield  {title} {\enquote {\bibinfo {title} {Enhanced protein adsorption
  due to charge regulation},}\ }\href@noop {} {\bibfield  {journal} {\bibinfo
  {journal} {Langmuir}\ }\textbf {\bibinfo {volume} {21}},\ \bibinfo {pages}
  {8385--8388} (\bibinfo {year} {2005})}\BibitemShut {NoStop}%
\bibitem [{\citenamefont {Nap}\ \emph {et~al.}(2014)\citenamefont {Nap},
  \citenamefont {Bo{\v{z}}i{\v{c}}}, \citenamefont {Szleifer},\ and\
  \citenamefont {Podgornik}}]{Nap2014}%
  \BibitemOpen
  \bibfield  {author} {\bibinfo {author} {\bibfnamefont {R.~J.}\ \bibnamefont
  {Nap}}, \bibinfo {author} {\bibfnamefont {A.~L.}\ \bibnamefont
  {Bo{\v{z}}i{\v{c}}}}, \bibinfo {author} {\bibfnamefont {I.}~\bibnamefont
  {Szleifer}}, \ and\ \bibinfo {author} {\bibfnamefont {R.}~\bibnamefont
  {Podgornik}},\ }\bibfield  {title} {\enquote {\bibinfo {title} {{The role of
  solution conditions in the bacteriophage pp7 capsid charge regulation}},}\
  }\href {\doibase 10.1016/j.bpj.2014.08.032} {\bibfield  {journal} {\bibinfo
  {journal} {Biophys. J.}\ }\textbf {\bibinfo {volume} {107}},\ \bibinfo
  {pages} {1970--1979} (\bibinfo {year} {2014})}\BibitemShut {NoStop}%
\bibitem [{\citenamefont {Lund}(2010)}]{Lundl}%
  \BibitemOpen
  \bibfield  {author} {\bibinfo {author} {\bibfnamefont {M.}~\bibnamefont
  {Lund}},\ }\bibfield  {title} {\enquote {\bibinfo {title} {Electrostatic
  chameleons in biological systems},}\ }\href {\doibase 10.1021/ja106480a}
  {\bibfield  {journal} {\bibinfo  {journal} {Journal of the American Chemical
  Society}\ }\textbf {\bibinfo {volume} {132}},\ \bibinfo {pages}
  {17337--17339} (\bibinfo {year} {2010})},\ \bibinfo {note} {pMID: 21086991},\
  \Eprint {http://arxiv.org/abs/https://doi.org/10.1021/ja106480a}
  {https://doi.org/10.1021/ja106480a} \BibitemShut {NoStop}%
\bibitem [{\citenamefont {Majee}\ \emph {et~al.}(2019)\citenamefont {Majee},
  \citenamefont {Bier}, \citenamefont {Blossey},\ and\ \citenamefont
  {Podgornik}}]{majee2019}%
  \BibitemOpen
  \bibfield  {author} {\bibinfo {author} {\bibfnamefont {A.}~\bibnamefont
  {Majee}}, \bibinfo {author} {\bibfnamefont {M.}~\bibnamefont {Bier}},
  \bibinfo {author} {\bibfnamefont {R.}~\bibnamefont {Blossey}}, \ and\
  \bibinfo {author} {\bibfnamefont {R.}~\bibnamefont {Podgornik}},\ }\bibfield
  {title} {\enquote {\bibinfo {title} {Charge regulation radically modifies
  electrostatics in membrane stacks},}\ }\href@noop {} {\bibfield  {journal}
  {\bibinfo  {journal} {Phys. Rev. E}\ }\textbf {\bibinfo {volume} {100}},\
  \bibinfo {pages} {050601} (\bibinfo {year} {2019})}\BibitemShut {NoStop}%
\bibitem [{\citenamefont {Netz}(2002)}]{netz2002}%
  \BibitemOpen
  \bibfield  {author} {\bibinfo {author} {\bibfnamefont {R.}~\bibnamefont
  {Netz}},\ }\bibfield  {title} {\enquote {\bibinfo {title} {Charge regulation
  of weak polyelectrolytes at low-and high-dielectric-constant substrates},}\
  }\href@noop {} {\bibfield  {journal} {\bibinfo  {journal} {J. Phys.: Condens.
  Matter}\ }\textbf {\bibinfo {volume} {15}},\ \bibinfo {pages} {S239}
  (\bibinfo {year} {2002})}\BibitemShut {NoStop}%
\bibitem [{\citenamefont {Israelachvili}(2011)}]{2011xxiii}%
  \BibitemOpen
  \bibfield  {author} {\bibinfo {author} {\bibfnamefont {J.~N.}\ \bibnamefont
  {Israelachvili}},\ }\bibfield  {title} {\enquote {\bibinfo {title} {Units,
  symbols, useful quantities and relations},}\ }in\ \href {\doibase
  https://doi.org/10.1016/B978-0-12-375182-9.10030-2} {\emph {\bibinfo
  {booktitle} {Intermolecular and Surface Forces (Third Edition)}}}\ (\bibinfo
  {publisher} {Academic Press},\ \bibinfo {address} {San Diego},\ \bibinfo
  {year} {2011})\ \bibinfo {edition} {third edition}\ ed.\BibitemShut {Stop}%
\bibitem [{\citenamefont {Behrens}\ \emph {et~al.}(2000)\citenamefont
  {Behrens}, \citenamefont {Christl}, \citenamefont {Emmerzael}, \citenamefont
  {Schurtenberger},\ and\ \citenamefont {Borkovec}}]{behrens2000charging}%
  \BibitemOpen
  \bibfield  {author} {\bibinfo {author} {\bibfnamefont {S.~H.}\ \bibnamefont
  {Behrens}}, \bibinfo {author} {\bibfnamefont {D.~I.}\ \bibnamefont
  {Christl}}, \bibinfo {author} {\bibfnamefont {R.}~\bibnamefont {Emmerzael}},
  \bibinfo {author} {\bibfnamefont {P.}~\bibnamefont {Schurtenberger}}, \ and\
  \bibinfo {author} {\bibfnamefont {M.}~\bibnamefont {Borkovec}},\ }\bibfield
  {title} {\enquote {\bibinfo {title} {Charging and aggregation properties of
  carboxyl latex particles: Experiments versus dlvo theory},}\ }\href@noop {}
  {\bibfield  {journal} {\bibinfo  {journal} {Langmuir}\ }\textbf {\bibinfo
  {volume} {16}},\ \bibinfo {pages} {2566--2575} (\bibinfo {year}
  {2000})}\BibitemShut {NoStop}%
\bibitem [{\citenamefont {Palaia}\ \emph {et~al.}(2022)\citenamefont {Palaia},
  \citenamefont {Goyal}, \citenamefont {Del~Gado}, \citenamefont {Šamaj},\
  and\ \citenamefont {Trizac}}]{Ivan11}%
  \BibitemOpen
  \bibfield  {author} {\bibinfo {author} {\bibfnamefont {I.}~\bibnamefont
  {Palaia}}, \bibinfo {author} {\bibfnamefont {A.}~\bibnamefont {Goyal}},
  \bibinfo {author} {\bibfnamefont {E.}~\bibnamefont {Del~Gado}}, \bibinfo
  {author} {\bibfnamefont {L.}~\bibnamefont {Šamaj}}, \ and\ \bibinfo {author}
  {\bibfnamefont {E.}~\bibnamefont {Trizac}},\ }\bibfield  {title} {\enquote
  {\bibinfo {title} {Like-charge attraction at the nanoscale: Ground-state
  correlations and water destructuring},}\ }\href {\doibase
  10.1021/acs.jpcb.2c00028} {\bibfield  {journal} {\bibinfo  {journal} {The
  Journal of Physical Chemistry B}\ }\textbf {\bibinfo {volume} {126}},\
  \bibinfo {pages} {3143--3149} (\bibinfo {year} {2022})},\ \bibinfo {note}
  {pMID: 35420420},\ \Eprint
  {http://arxiv.org/abs/https://doi.org/10.1021/acs.jpcb.2c00028}
  {https://doi.org/10.1021/acs.jpcb.2c00028} \BibitemShut {NoStop}%
\bibitem [{\citenamefont {Lund}\ and\ \citenamefont
  {J{\"{o}}nsson}(2013)}]{Lund2013}%
  \BibitemOpen
  \bibfield  {author} {\bibinfo {author} {\bibfnamefont {M.}~\bibnamefont
  {Lund}}\ and\ \bibinfo {author} {\bibfnamefont {B.}~\bibnamefont
  {J{\"{o}}nsson}},\ }\bibfield  {title} {\enquote {\bibinfo {title} {{Charge
  regulation in biomolecular solution}},}\ }\href {\doibase
  10.1017/S003358351300005X} {\bibfield  {journal} {\bibinfo  {journal} {Q.
  Rev. Biophys.}\ }\textbf {\bibinfo {volume} {46}},\ \bibinfo {pages}
  {265--281} (\bibinfo {year} {2013})}\BibitemShut {NoStop}%
\bibitem [{\citenamefont {Lund}\ and\ \citenamefont
  {J{\"o}nsson}(2005)}]{lund2005charge}%
  \BibitemOpen
  \bibfield  {author} {\bibinfo {author} {\bibfnamefont {M.}~\bibnamefont
  {Lund}}\ and\ \bibinfo {author} {\bibfnamefont {B.}~\bibnamefont
  {J{\"o}nsson}},\ }\bibfield  {title} {\enquote {\bibinfo {title} {On the
  charge regulation of proteins},}\ }\href@noop {} {\bibfield  {journal}
  {\bibinfo  {journal} {Biochemistry}\ }\textbf {\bibinfo {volume} {44}},\
  \bibinfo {pages} {5722--5727} (\bibinfo {year} {2005})}\BibitemShut {NoStop}%
\bibitem [{\citenamefont {Landsgesell}\ \emph {et~al.}(2019)\citenamefont
  {Landsgesell}, \citenamefont {Nov{\'a}}, \citenamefont {Rud}, \citenamefont
  {Uhl{\'\i}k}, \citenamefont {Sean}, \citenamefont {Hebbeker}, \citenamefont
  {Holm},\ and\ \citenamefont {Ko{\v{s}}ovan}}]{landsgesell2019simulations}%
  \BibitemOpen
  \bibfield  {author} {\bibinfo {author} {\bibfnamefont {J.}~\bibnamefont
  {Landsgesell}}, \bibinfo {author} {\bibfnamefont {L.}~\bibnamefont
  {Nov{\'a}}}, \bibinfo {author} {\bibfnamefont {O.}~\bibnamefont {Rud}},
  \bibinfo {author} {\bibfnamefont {F.}~\bibnamefont {Uhl{\'\i}k}}, \bibinfo
  {author} {\bibfnamefont {D.}~\bibnamefont {Sean}}, \bibinfo {author}
  {\bibfnamefont {P.}~\bibnamefont {Hebbeker}}, \bibinfo {author}
  {\bibfnamefont {C.}~\bibnamefont {Holm}}, \ and\ \bibinfo {author}
  {\bibfnamefont {P.}~\bibnamefont {Ko{\v{s}}ovan}},\ }\bibfield  {title}
  {\enquote {\bibinfo {title} {Simulations of ionization equilibria in weak
  polyelectrolyte solutions and gels},}\ }\href@noop {} {\bibfield  {journal}
  {\bibinfo  {journal} {Soft Matter}\ }\textbf {\bibinfo {volume} {15}},\
  \bibinfo {pages} {1155--1185} (\bibinfo {year} {2019})}\BibitemShut {NoStop}%
\bibitem [{\citenamefont {P{\'e}rez-Ch{\'a}vez}, \citenamefont {Albesa},\ and\
  \citenamefont {Longo}(2021)}]{perez2021thermodynamic}%
  \BibitemOpen
  \bibfield  {author} {\bibinfo {author} {\bibfnamefont {N.~A.}\ \bibnamefont
  {P{\'e}rez-Ch{\'a}vez}}, \bibinfo {author} {\bibfnamefont {A.~G.}\
  \bibnamefont {Albesa}}, \ and\ \bibinfo {author} {\bibfnamefont {G.~S.}\
  \bibnamefont {Longo}},\ }\bibfield  {title} {\enquote {\bibinfo {title}
  {Thermodynamic theory of multiresponsive microgel swelling},}\ }\href@noop {}
  {\bibfield  {journal} {\bibinfo  {journal} {Macromolecules}\ }\textbf
  {\bibinfo {volume} {54}},\ \bibinfo {pages} {2936--2947} (\bibinfo {year}
  {2021})}\BibitemShut {NoStop}%
\bibitem [{\citenamefont {Burak}\ and\ \citenamefont {Netz}(2004)}]{Burak}%
  \BibitemOpen
  \bibfield  {author} {\bibinfo {author} {\bibfnamefont {Y.}~\bibnamefont
  {Burak}}\ and\ \bibinfo {author} {\bibfnamefont {R.~R.}\ \bibnamefont
  {Netz}},\ }\bibfield  {title} {\enquote {\bibinfo {title} {Charge regulation
  of interacting weak polyelectrolytes},}\ }\href@noop {} {\bibfield  {journal}
  {\bibinfo  {journal} {J. Phys. Chem. B}\ }\textbf {\bibinfo {volume} {108}},\
  \bibinfo {pages} {4840--4849} (\bibinfo {year} {2004})}\BibitemShut {NoStop}%
\bibitem [{\citenamefont {Hiemstra}, \citenamefont {Van~Riemsdijk},\ and\
  \citenamefont {Bruggenwert}(1987)}]{hiemstra1987proton}%
  \BibitemOpen
  \bibfield  {author} {\bibinfo {author} {\bibfnamefont {T.}~\bibnamefont
  {Hiemstra}}, \bibinfo {author} {\bibfnamefont {W.}~\bibnamefont
  {Van~Riemsdijk}}, \ and\ \bibinfo {author} {\bibfnamefont {M.}~\bibnamefont
  {Bruggenwert}},\ }\bibfield  {title} {\enquote {\bibinfo {title} {Proton
  adsorption mechanism at the gibbsite and aluminium oxide solid/solution
  interface.}}\ }\href@noop {} {\bibfield  {journal} {\bibinfo  {journal}
  {Netherlands journal of agricultural science}\ }\textbf {\bibinfo {volume}
  {35}},\ \bibinfo {pages} {281--293} (\bibinfo {year} {1987})}\BibitemShut
  {NoStop}%
\bibitem [{\citenamefont {Hiemstra}\ and\ \citenamefont
  {Van~Riemsdijk}(1991)}]{hiemstra1991physical}%
  \BibitemOpen
  \bibfield  {author} {\bibinfo {author} {\bibfnamefont {T.}~\bibnamefont
  {Hiemstra}}\ and\ \bibinfo {author} {\bibfnamefont {W.}~\bibnamefont
  {Van~Riemsdijk}},\ }\bibfield  {title} {\enquote {\bibinfo {title} {Physical
  chemical interpretation of primary charging behaviour of metal (hydr)
  oxides},}\ }\href@noop {} {\bibfield  {journal} {\bibinfo  {journal}
  {Colloids and surfaces}\ }\textbf {\bibinfo {volume} {59}},\ \bibinfo {pages}
  {7--25} (\bibinfo {year} {1991})}\BibitemShut {NoStop}%
\bibitem [{\citenamefont {Bakhshandeh}, \citenamefont {Dos~Santos},\ and\
  \citenamefont {Levin}(2020)}]{bakhshandeh2020interaction}%
  \BibitemOpen
  \bibfield  {author} {\bibinfo {author} {\bibfnamefont {A.}~\bibnamefont
  {Bakhshandeh}}, \bibinfo {author} {\bibfnamefont {A.~P.}\ \bibnamefont
  {Dos~Santos}}, \ and\ \bibinfo {author} {\bibfnamefont {Y.}~\bibnamefont
  {Levin}},\ }\bibfield  {title} {\enquote {\bibinfo {title} {Interaction
  between charge-regulated metal nanoparticles in an electrolyte solution},}\
  }\href@noop {} {\bibfield  {journal} {\bibinfo  {journal} {The Journal of
  Physical Chemistry B}\ }\textbf {\bibinfo {volume} {124}},\ \bibinfo {pages}
  {11762--11770} (\bibinfo {year} {2020})}\BibitemShut {NoStop}%
\bibitem [{\citenamefont {Bakhshandeh}, \citenamefont {Frydel},\ and\
  \citenamefont {Levin}(2022{\natexlab{a}})}]{bkh2022}%
  \BibitemOpen
  \bibfield  {author} {\bibinfo {author} {\bibfnamefont {A.}~\bibnamefont
  {Bakhshandeh}}, \bibinfo {author} {\bibfnamefont {D.}~\bibnamefont {Frydel}},
  \ and\ \bibinfo {author} {\bibfnamefont {Y.}~\bibnamefont {Levin}},\
  }\bibfield  {title} {\enquote {\bibinfo {title} {Theory of charge regulation
  of colloidal particles in electrolyte solutions},}\ }\href@noop {} {\bibfield
   {journal} {\bibinfo  {journal} {Langmuir}\ } (\bibinfo {year}
  {2022}{\natexlab{a}})}\BibitemShut {NoStop}%
\bibitem [{\citenamefont {Levin}(2002)}]{Le02}%
  \BibitemOpen
  \bibfield  {author} {\bibinfo {author} {\bibfnamefont {Y.}~\bibnamefont
  {Levin}},\ }\bibfield  {title} {\enquote {\bibinfo {title} {Electrostatic
  correlations: from plasma to biology},}\ }\href@noop {} {\bibfield  {journal}
  {\bibinfo  {journal} {Rep. Prog. Phys.}\ }\textbf {\bibinfo {volume} {65}},\
  \bibinfo {pages} {1577} (\bibinfo {year} {2002})}\BibitemShut {NoStop}%
\bibitem [{\citenamefont {Levin}\ and\ \citenamefont
  {Fisher}(1996)}]{levin1996criticality}%
  \BibitemOpen
  \bibfield  {author} {\bibinfo {author} {\bibfnamefont {Y.}~\bibnamefont
  {Levin}}\ and\ \bibinfo {author} {\bibfnamefont {M.~E.}\ \bibnamefont
  {Fisher}},\ }\bibfield  {title} {\enquote {\bibinfo {title} {Criticality in
  the hard-sphere ionic fluid},}\ }\href@noop {} {\bibfield  {journal}
  {\bibinfo  {journal} {Physica A: Statistical Mechanics and its Applications}\
  }\textbf {\bibinfo {volume} {225}},\ \bibinfo {pages} {164--220} (\bibinfo
  {year} {1996})}\BibitemShut {NoStop}%
\bibitem [{\citenamefont {Labbez}\ and\ \citenamefont
  {J{\"o}nsson}(2006)}]{labbez2006new}%
  \BibitemOpen
  \bibfield  {author} {\bibinfo {author} {\bibfnamefont {C.}~\bibnamefont
  {Labbez}}\ and\ \bibinfo {author} {\bibfnamefont {B.}~\bibnamefont
  {J{\"o}nsson}},\ }\href@noop {} {\emph {\bibinfo {title} {A new Monte Carlo
  method for the titration of molecules and minerals}}}\ (\bibinfo  {publisher}
  {Springer},\ \bibinfo {year} {2006})\ pp.\ \bibinfo {pages}
  {66--72}\BibitemShut {NoStop}%
\bibitem [{\citenamefont {Labbez}\ \emph {et~al.}(2009)\citenamefont {Labbez},
  \citenamefont {J\"{o}nsson}, \citenamefont {Skarba},\ and\ \citenamefont
  {Borkovec}}]{Labbezl}%
  \BibitemOpen
  \bibfield  {author} {\bibinfo {author} {\bibfnamefont {C.}~\bibnamefont
  {Labbez}}, \bibinfo {author} {\bibfnamefont {B.}~\bibnamefont {J\"{o}nsson}},
  \bibinfo {author} {\bibfnamefont {M.}~\bibnamefont {Skarba}}, \ and\ \bibinfo
  {author} {\bibfnamefont {M.}~\bibnamefont {Borkovec}},\ }\bibfield  {title}
  {\enquote {\bibinfo {title} {Ion−ion correlation and charge reversal at
  titrating solid interfaces},}\ }\href {\doibase 10.1021/la900853e} {\bibfield
   {journal} {\bibinfo  {journal} {Langmuir}\ }\textbf {\bibinfo {volume}
  {25}},\ \bibinfo {pages} {7209--7213} (\bibinfo {year} {2009})}\BibitemShut
  {NoStop}%
\bibitem [{\citenamefont {Levin}\ and\ \citenamefont
  {Bakhshandeh}(2023)}]{Levincomment}%
  \BibitemOpen
  \bibfield  {author} {\bibinfo {author} {\bibfnamefont {Y.}~\bibnamefont
  {Levin}}\ and\ \bibinfo {author} {\bibfnamefont {A.}~\bibnamefont
  {Bakhshandeh}},\ }\bibfield  {title} {\enquote {\bibinfo {title} {Comment on
  “simulations of ionization equilibria in weak polyelectrolyte solutions and
  gels” by j. landsgesell{,} l. nová{,} o. rud{,} f. uhlík{,} d. sean{,} p.
  hebbeker{,} c. holm and p. košovan{,} soft matter{,} 2019{,} 15{,}
  1155–1185},}\ }\href@noop {} {\bibfield  {journal} {\bibinfo  {journal}
  {Soft Matter}\ }\textbf {\bibinfo {volume} {19}},\ \bibinfo {pages}
  {3519--3521} (\bibinfo {year} {2023})}\BibitemShut {NoStop}%
\bibitem [{\citenamefont {Smith}\ and\ \citenamefont
  {Triska}(1994)}]{smith1994reaction}%
  \BibitemOpen
  \bibfield  {author} {\bibinfo {author} {\bibfnamefont {W.}~\bibnamefont
  {Smith}}\ and\ \bibinfo {author} {\bibfnamefont {B.}~\bibnamefont {Triska}},\
  }\bibfield  {title} {\enquote {\bibinfo {title} {The reaction ensemble method
  for the computer simulation of chemical and phase equilibria. i. theory and
  basic examples},}\ }\href@noop {} {\bibfield  {journal} {\bibinfo  {journal}
  {The Journal of chemical physics}\ }\textbf {\bibinfo {volume} {100}},\
  \bibinfo {pages} {3019--3027} (\bibinfo {year} {1994})}\BibitemShut {NoStop}%
\bibitem [{\citenamefont {Johnson}, \citenamefont {Panagiotopoulos},\ and\
  \citenamefont {Gubbins}(1994)}]{johnson1994reactive}%
  \BibitemOpen
  \bibfield  {author} {\bibinfo {author} {\bibfnamefont {J.~K.}\ \bibnamefont
  {Johnson}}, \bibinfo {author} {\bibfnamefont {A.~Z.}\ \bibnamefont
  {Panagiotopoulos}}, \ and\ \bibinfo {author} {\bibfnamefont {K.~E.}\
  \bibnamefont {Gubbins}},\ }\bibfield  {title} {\enquote {\bibinfo {title}
  {Reactive canonical monte carlo: a new simulation technique for reacting or
  associating fluids},}\ }\href@noop {} {\bibfield  {journal} {\bibinfo
  {journal} {Molecular Physics}\ }\textbf {\bibinfo {volume} {81}},\ \bibinfo
  {pages} {717--733} (\bibinfo {year} {1994})}\BibitemShut {NoStop}%
\bibitem [{\citenamefont {Leung}\ \emph {et~al.}(2013)\citenamefont {Leung},
  \citenamefont {Palmer}, \citenamefont {Kewalramani}, \citenamefont {Qiao},
  \citenamefont {Stupp}, \citenamefont {Olvera de~la Cruz},\ and\ \citenamefont
  {Bedzyk}}]{monica2}%
  \BibitemOpen
  \bibfield  {author} {\bibinfo {author} {\bibfnamefont {C.-Y.}\ \bibnamefont
  {Leung}}, \bibinfo {author} {\bibfnamefont {L.~C.}\ \bibnamefont {Palmer}},
  \bibinfo {author} {\bibfnamefont {S.}~\bibnamefont {Kewalramani}}, \bibinfo
  {author} {\bibfnamefont {B.}~\bibnamefont {Qiao}}, \bibinfo {author}
  {\bibfnamefont {S.~I.}\ \bibnamefont {Stupp}}, \bibinfo {author}
  {\bibfnamefont {M.}~\bibnamefont {Olvera de~la Cruz}}, \ and\ \bibinfo
  {author} {\bibfnamefont {M.~J.}\ \bibnamefont {Bedzyk}},\ }\bibfield  {title}
  {\enquote {\bibinfo {title} {Crystalline polymorphism induced by charge
  regulation in ionic membranes},}\ }\href {\doibase 10.1073/pnas.1316150110}
  {\bibfield  {journal} {\bibinfo  {journal} {Proceedings of the National
  Academy of Sciences}\ }\textbf {\bibinfo {volume} {110}},\ \bibinfo {pages}
  {16309--16314} (\bibinfo {year} {2013})}\BibitemShut {NoStop}%
\bibitem [{\citenamefont {Ong}, \citenamefont {Gallegos},\ and\ \citenamefont
  {Wu}(2020)}]{Ong}%
  \BibitemOpen
  \bibfield  {author} {\bibinfo {author} {\bibfnamefont {G.~M.}\ \bibnamefont
  {Ong}}, \bibinfo {author} {\bibfnamefont {A.}~\bibnamefont {Gallegos}}, \
  and\ \bibinfo {author} {\bibfnamefont {J.}~\bibnamefont {Wu}},\ }\bibfield
  {title} {\enquote {\bibinfo {title} {Modeling surface charge regulation of
  colloidal particles in aqueous solutions},}\ }\href@noop {} {\bibfield
  {journal} {\bibinfo  {journal} {Langmuir}\ }\textbf {\bibinfo {volume}
  {36}},\ \bibinfo {pages} {11918--11928} (\bibinfo {year} {2020})}\BibitemShut
  {NoStop}%
\bibitem [{\citenamefont {Wennerstr{\"o}m}, \citenamefont {J{\"o}nsson},\ and\
  \citenamefont {Linse}(1982)}]{wenncell}%
  \BibitemOpen
  \bibfield  {author} {\bibinfo {author} {\bibfnamefont {H.}~\bibnamefont
  {Wennerstr{\"o}m}}, \bibinfo {author} {\bibfnamefont {B.}~\bibnamefont
  {J{\"o}nsson}}, \ and\ \bibinfo {author} {\bibfnamefont {P.}~\bibnamefont
  {Linse}},\ }\bibfield  {title} {\enquote {\bibinfo {title} {The cell model
  for polyelectrolyte systems. exact statistical mechanical relations, monte
  carlo simulations, and the poisson--boltzmann approximation},}\ }\href@noop
  {} {\bibfield  {journal} {\bibinfo  {journal} {The Journal of Chemical
  Physics}\ }\textbf {\bibinfo {volume} {76}},\ \bibinfo {pages} {4665--4670}
  (\bibinfo {year} {1982})}\BibitemShut {NoStop}%
\bibitem [{\citenamefont {Landsgesell}\ \emph {et~al.}(2020)\citenamefont
  {Landsgesell}, \citenamefont {Hebbeker}, \citenamefont {Rud}, \citenamefont
  {Lunkad}, \citenamefont {Ko\v{s}ovan},\ and\ \citenamefont {Holm}}]{lan}%
  \BibitemOpen
  \bibfield  {author} {\bibinfo {author} {\bibfnamefont {J.}~\bibnamefont
  {Landsgesell}}, \bibinfo {author} {\bibfnamefont {P.}~\bibnamefont
  {Hebbeker}}, \bibinfo {author} {\bibfnamefont {O.}~\bibnamefont {Rud}},
  \bibinfo {author} {\bibfnamefont {R.}~\bibnamefont {Lunkad}}, \bibinfo
  {author} {\bibfnamefont {P.}~\bibnamefont {Ko\v{s}ovan}}, \ and\ \bibinfo
  {author} {\bibfnamefont {C.}~\bibnamefont {Holm}},\ }\bibfield  {title}
  {\enquote {\bibinfo {title} {Grand-reaction method for simulations of
  ionization equilibria coupled to ion partitioning},}\ }\href@noop {}
  {\bibfield  {journal} {\bibinfo  {journal} {Macromolecules}\ }\textbf
  {\bibinfo {volume} {53}},\ \bibinfo {pages} {3007--3020} (\bibinfo {year}
  {2020})}\BibitemShut {NoStop}%
\bibitem [{\citenamefont {Curk}, \citenamefont {Yuan},\ and\ \citenamefont
  {Luijten}(2022)}]{curk1}%
  \BibitemOpen
  \bibfield  {author} {\bibinfo {author} {\bibfnamefont {T.}~\bibnamefont
  {Curk}}, \bibinfo {author} {\bibfnamefont {J.}~\bibnamefont {Yuan}}, \ and\
  \bibinfo {author} {\bibfnamefont {E.}~\bibnamefont {Luijten}},\ }\bibfield
  {title} {\enquote {\bibinfo {title} {Accelerated simulation method for charge
  regulation effects},}\ }\href@noop {} {\bibfield  {journal} {\bibinfo
  {journal} {The Journal of Chemical Physics}\ }\textbf {\bibinfo {volume}
  {156}} (\bibinfo {year} {2022})}\BibitemShut {NoStop}%
\bibitem [{\citenamefont {Bollhorst}, \citenamefont {Rezwan},\ and\
  \citenamefont {Maas}(2017)}]{C6CS00632A}%
  \BibitemOpen
  \bibfield  {author} {\bibinfo {author} {\bibfnamefont {T.}~\bibnamefont
  {Bollhorst}}, \bibinfo {author} {\bibfnamefont {K.}~\bibnamefont {Rezwan}}, \
  and\ \bibinfo {author} {\bibfnamefont {M.}~\bibnamefont {Maas}},\ }\bibfield
  {title} {\enquote {\bibinfo {title} {Colloidal capsules: nano- and
  microcapsules with colloidal particle shells},}\ }\href {\doibase
  10.1039/C6CS00632A} {\bibfield  {journal} {\bibinfo  {journal} {Chem. Soc.
  Rev.}\ }\textbf {\bibinfo {volume} {46}},\ \bibinfo {pages} {2091--2126}
  (\bibinfo {year} {2017})}\BibitemShut {NoStop}%
\bibitem [{\citenamefont {Donnan}(1911)}]{donnan1911theorie}%
  \BibitemOpen
  \bibfield  {author} {\bibinfo {author} {\bibfnamefont {F.~G.}\ \bibnamefont
  {Donnan}},\ }\bibfield  {title} {\enquote {\bibinfo {title} {Theorie der
  membrangleichgewichte und membranpotentiale bei vorhandensein von nicht
  dialysierenden elektrolyten. ein beitrag zur physikalisch-chemischen
  physiologie.}}\ }\href@noop {} {\bibfield  {journal} {\bibinfo  {journal}
  {Zeitschrift f{\"u}r Elektrochemie und angewandte physikalische Chemie}\
  }\textbf {\bibinfo {volume} {17}},\ \bibinfo {pages} {572--581} (\bibinfo
  {year} {1911})}\BibitemShut {NoStop}%
\bibitem [{\citenamefont {Barr}\ and\ \citenamefont
  {Panagiotopoulos}(2012)}]{barr2012grand}%
  \BibitemOpen
  \bibfield  {author} {\bibinfo {author} {\bibfnamefont {S.}~\bibnamefont
  {Barr}}\ and\ \bibinfo {author} {\bibfnamefont {A.}~\bibnamefont
  {Panagiotopoulos}},\ }\bibfield  {title} {\enquote {\bibinfo {title}
  {Grand-canonical monte carlo method for donnan equilibria},}\ }\href@noop {}
  {\bibfield  {journal} {\bibinfo  {journal} {Physical Review E}\ }\textbf
  {\bibinfo {volume} {86}},\ \bibinfo {pages} {016703} (\bibinfo {year}
  {2012})}\BibitemShut {NoStop}%
\bibitem [{\citenamefont {van Roij}(2003)}]{van2003defying}%
  \BibitemOpen
  \bibfield  {author} {\bibinfo {author} {\bibfnamefont {R.}~\bibnamefont {van
  Roij}},\ }\bibfield  {title} {\enquote {\bibinfo {title} {Defying gravity
  with entropy and electrostatics: sedimentation of charged colloids},}\
  }\href@noop {} {\bibfield  {journal} {\bibinfo  {journal} {Journal of
  Physics: Condensed Matter}\ }\textbf {\bibinfo {volume} {15}},\ \bibinfo
  {pages} {S3569} (\bibinfo {year} {2003})}\BibitemShut {NoStop}%
\bibitem [{\citenamefont {Philipse}(2004)}]{philipse2004remarks}%
  \BibitemOpen
  \bibfield  {author} {\bibinfo {author} {\bibfnamefont {A.~P.}\ \bibnamefont
  {Philipse}},\ }\bibfield  {title} {\enquote {\bibinfo {title} {Remarks on the
  donnan condenser in the sedimentation--diffusion equilibrium of charged
  colloids},}\ }\href@noop {} {\bibfield  {journal} {\bibinfo  {journal}
  {Journal of Physics: Condensed Matter}\ }\textbf {\bibinfo {volume} {16}},\
  \bibinfo {pages} {S4051} (\bibinfo {year} {2004})}\BibitemShut {NoStop}%
\bibitem [{\citenamefont {Warren}(2004)}]{warren2004electrifying}%
  \BibitemOpen
  \bibfield  {author} {\bibinfo {author} {\bibfnamefont {P.}~\bibnamefont
  {Warren}},\ }\bibfield  {title} {\enquote {\bibinfo {title} {Electrifying
  effects in colloids},}\ }\href@noop {} {\bibfield  {journal} {\bibinfo
  {journal} {Nature}\ }\textbf {\bibinfo {volume} {429}},\ \bibinfo {pages}
  {822--822} (\bibinfo {year} {2004})}\BibitemShut {NoStop}%
\bibitem [{\citenamefont {Eckert}, \citenamefont {Schmidt},\ and\ \citenamefont
  {de~las Heras}(2022)}]{eckert2022sedimentation}%
  \BibitemOpen
  \bibfield  {author} {\bibinfo {author} {\bibfnamefont {T.}~\bibnamefont
  {Eckert}}, \bibinfo {author} {\bibfnamefont {M.}~\bibnamefont {Schmidt}}, \
  and\ \bibinfo {author} {\bibfnamefont {D.}~\bibnamefont {de~las Heras}},\
  }\bibfield  {title} {\enquote {\bibinfo {title} {Sedimentation of colloidal
  plate-sphere mixtures and inference of particle characteristics from stacking
  sequences},}\ }\href@noop {} {\bibfield  {journal} {\bibinfo  {journal}
  {Physical Review Research}\ }\textbf {\bibinfo {volume} {4}},\ \bibinfo
  {pages} {013189} (\bibinfo {year} {2022})}\BibitemShut {NoStop}%
\bibitem [{\citenamefont {Sullivan}\ \emph {et~al.}(2002)\citenamefont
  {Sullivan}, \citenamefont {Zhao}, \citenamefont {Harrison}, \citenamefont
  {Austin}, \citenamefont {Megens}, \citenamefont {Hollingsworth},
  \citenamefont {Russel}, \citenamefont {Cheng}, \citenamefont {Mason},\ and\
  \citenamefont {Chaikin}}]{sullivan2002control}%
  \BibitemOpen
  \bibfield  {author} {\bibinfo {author} {\bibfnamefont {M.}~\bibnamefont
  {Sullivan}}, \bibinfo {author} {\bibfnamefont {K.}~\bibnamefont {Zhao}},
  \bibinfo {author} {\bibfnamefont {C.}~\bibnamefont {Harrison}}, \bibinfo
  {author} {\bibfnamefont {R.~H.}\ \bibnamefont {Austin}}, \bibinfo {author}
  {\bibfnamefont {M.}~\bibnamefont {Megens}}, \bibinfo {author} {\bibfnamefont
  {A.}~\bibnamefont {Hollingsworth}}, \bibinfo {author} {\bibfnamefont {W.~B.}\
  \bibnamefont {Russel}}, \bibinfo {author} {\bibfnamefont {Z.}~\bibnamefont
  {Cheng}}, \bibinfo {author} {\bibfnamefont {T.}~\bibnamefont {Mason}}, \ and\
  \bibinfo {author} {\bibfnamefont {P.}~\bibnamefont {Chaikin}},\ }\bibfield
  {title} {\enquote {\bibinfo {title} {Control of colloids with gravity,
  temperature gradients, and electric fields},}\ }\href@noop {} {\bibfield
  {journal} {\bibinfo  {journal} {Journal of Physics: Condensed Matter}\
  }\textbf {\bibinfo {volume} {15}},\ \bibinfo {pages} {S11} (\bibinfo {year}
  {2002})}\BibitemShut {NoStop}%
\bibitem [{\citenamefont {Smith}(1981)}]{smith1981electrostatic}%
  \BibitemOpen
  \bibfield  {author} {\bibinfo {author} {\bibfnamefont {E.~R.}\ \bibnamefont
  {Smith}},\ }\bibfield  {title} {\enquote {\bibinfo {title} {Electrostatic
  energy in ionic crystals},}\ }\href@noop {} {\bibfield  {journal} {\bibinfo
  {journal} {Proceedings of the Royal Society of London. A. Mathematical and
  Physical Sciences}\ }\textbf {\bibinfo {volume} {375}},\ \bibinfo {pages}
  {475--505} (\bibinfo {year} {1981})}\BibitemShut {NoStop}%
\bibitem [{\citenamefont {Frenkel}\ and\ \citenamefont
  {Smit}(2001)}]{frenkel2001understanding}%
  \BibitemOpen
  \bibfield  {author} {\bibinfo {author} {\bibfnamefont {D.}~\bibnamefont
  {Frenkel}}\ and\ \bibinfo {author} {\bibfnamefont {B.}~\bibnamefont {Smit}},\
  }\href@noop {} {\emph {\bibinfo {title} {Understanding molecular simulation:
  from algorithms to applications}}},\ Vol.~\bibinfo {volume} {1}\ (\bibinfo
  {publisher} {Elsevier},\ \bibinfo {year} {2001})\BibitemShut {NoStop}%
\bibitem [{\citenamefont {Allen}\ and\ \citenamefont
  {Tildesley}(2017)}]{allen2017computer}%
  \BibitemOpen
  \bibfield  {author} {\bibinfo {author} {\bibfnamefont {M.~P.}\ \bibnamefont
  {Allen}}\ and\ \bibinfo {author} {\bibfnamefont {D.~J.}\ \bibnamefont
  {Tildesley}},\ }\href@noop {} {\emph {\bibinfo {title} {Computer simulation
  of liquids}}}\ (\bibinfo  {publisher} {Oxford university press},\ \bibinfo
  {year} {2017})\BibitemShut {NoStop}%
\bibitem [{\citenamefont {Hu}(2014)}]{Zhonghan}%
  \BibitemOpen
  \bibfield  {author} {\bibinfo {author} {\bibfnamefont {Z.}~\bibnamefont
  {Hu}},\ }\bibfield  {title} {\enquote {\bibinfo {title} {Infinite boundary
  terms of ewald sums and pairwise interactions for electrostatics in bulk and
  at interfaces},}\ }\href {\doibase 10.1021/ct500704m} {\bibfield  {journal}
  {\bibinfo  {journal} {Journal of Chemical Theory and Computation}\ }\textbf
  {\bibinfo {volume} {10}},\ \bibinfo {pages} {5254--5264} (\bibinfo {year}
  {2014})},\ \bibinfo {note} {pMID: 26583209},\ \Eprint
  {http://arxiv.org/abs/https://doi.org/10.1021/ct500704m}
  {https://doi.org/10.1021/ct500704m} \BibitemShut {NoStop}%
\bibitem [{\citenamefont {dos Santos}, \citenamefont {Girotto},\ and\
  \citenamefont {Levin}(2016)}]{dos2016simulations}%
  \BibitemOpen
  \bibfield  {author} {\bibinfo {author} {\bibfnamefont {A.~P.}\ \bibnamefont
  {dos Santos}}, \bibinfo {author} {\bibfnamefont {M.}~\bibnamefont {Girotto}},
  \ and\ \bibinfo {author} {\bibfnamefont {Y.}~\bibnamefont {Levin}},\
  }\bibfield  {title} {\enquote {\bibinfo {title} {Simulations of coulomb
  systems with slab geometry using an efficient 3d ewald summation method},}\
  }\href@noop {} {\bibfield  {journal} {\bibinfo  {journal} {The Journal of
  chemical physics}\ }\textbf {\bibinfo {volume} {144}},\ \bibinfo {pages}
  {144103} (\bibinfo {year} {2016})}\BibitemShut {NoStop}%
\bibitem [{\citenamefont {Yi}, \citenamefont {Pan},\ and\ \citenamefont
  {Hu}(2017)}]{Shasha}%
  \BibitemOpen
  \bibfield  {author} {\bibinfo {author} {\bibfnamefont {S.}~\bibnamefont
  {Yi}}, \bibinfo {author} {\bibfnamefont {C.}~\bibnamefont {Pan}}, \ and\
  \bibinfo {author} {\bibfnamefont {Z.}~\bibnamefont {Hu}},\ }\bibfield
  {title} {\enquote {\bibinfo {title} {Note: A pairwise form of the ewald sum
  for non-neutral systems},}\ }\href {\doibase 10.1063/1.4998320} {\bibfield
  {journal} {\bibinfo  {journal} {The Journal of Chemical Physics}\ }\textbf
  {\bibinfo {volume} {147}},\ \bibinfo {pages} {126101} (\bibinfo {year}
  {2017})},\ \Eprint {http://arxiv.org/abs/https://doi.org/10.1063/1.4998320}
  {https://doi.org/10.1063/1.4998320} \BibitemShut {NoStop}%
\bibitem [{\citenamefont {Euwema}\ and\ \citenamefont
  {Surratt}(1975)}]{euwema1975absolute}%
  \BibitemOpen
  \bibfield  {author} {\bibinfo {author} {\bibfnamefont {R.}~\bibnamefont
  {Euwema}}\ and\ \bibinfo {author} {\bibfnamefont {G.}~\bibnamefont
  {Surratt}},\ }\bibfield  {title} {\enquote {\bibinfo {title} {The absolute
  positions of calculated energy bands},}\ }\href@noop {} {\bibfield  {journal}
  {\bibinfo  {journal} {Journal of Physics and Chemistry of Solids}\ }\textbf
  {\bibinfo {volume} {36}},\ \bibinfo {pages} {67--71} (\bibinfo {year}
  {1975})}\BibitemShut {NoStop}%
\bibitem [{\citenamefont {De~Leeuw}\ and\ \citenamefont
  {Perram}(1981)}]{de1981computer}%
  \BibitemOpen
  \bibfield  {author} {\bibinfo {author} {\bibfnamefont {S.}~\bibnamefont
  {De~Leeuw}}\ and\ \bibinfo {author} {\bibfnamefont {J.}~\bibnamefont
  {Perram}},\ }\bibfield  {title} {\enquote {\bibinfo {title} {Computer
  simulation of ionic systems. influence of boundary conditions},}\ }\href@noop
  {} {\bibfield  {journal} {\bibinfo  {journal} {Physica A: Statistical
  Mechanics and its Applications}\ }\textbf {\bibinfo {volume} {107}},\
  \bibinfo {pages} {179--189} (\bibinfo {year} {1981})}\BibitemShut {NoStop}%
\bibitem [{\citenamefont {H{\o}ye}\ and\ \citenamefont
  {Lomba}(1988)}]{ho1988mean}%
  \BibitemOpen
  \bibfield  {author} {\bibinfo {author} {\bibfnamefont {J.~S.}\ \bibnamefont
  {H{\o}ye}}\ and\ \bibinfo {author} {\bibfnamefont {E.}~\bibnamefont
  {Lomba}},\ }\bibfield  {title} {\enquote {\bibinfo {title} {Mean spherical
  approximation (msa) for a simple model of electrolytes. i. theoretical
  foundations and thermodynamics},}\ }\href@noop {} {\bibfield  {journal}
  {\bibinfo  {journal} {The Journal of chemical physics}\ }\textbf {\bibinfo
  {volume} {88}},\ \bibinfo {pages} {5790--5797} (\bibinfo {year}
  {1988})}\BibitemShut {NoStop}%
\bibitem [{\citenamefont {Carnahan}\ and\ \citenamefont
  {Starling}(1969)}]{carnahan1969equation}%
  \BibitemOpen
  \bibfield  {author} {\bibinfo {author} {\bibfnamefont {N.~F.}\ \bibnamefont
  {Carnahan}}\ and\ \bibinfo {author} {\bibfnamefont {K.~E.}\ \bibnamefont
  {Starling}},\ }\bibfield  {title} {\enquote {\bibinfo {title} {Equation of
  state for nonattracting rigid spheres},}\ }\href@noop {} {\bibfield
  {journal} {\bibinfo  {journal} {The Journal of chemical physics}\ }\textbf
  {\bibinfo {volume} {51}},\ \bibinfo {pages} {635--636} (\bibinfo {year}
  {1969})}\BibitemShut {NoStop}%
\bibitem [{\citenamefont {Bakhshandeh}, \citenamefont {Frydel},\ and\
  \citenamefont {Levin}(2022{\natexlab{b}})}]{reactive}%
  \BibitemOpen
  \bibfield  {author} {\bibinfo {author} {\bibfnamefont {A.}~\bibnamefont
  {Bakhshandeh}}, \bibinfo {author} {\bibfnamefont {D.}~\bibnamefont {Frydel}},
  \ and\ \bibinfo {author} {\bibfnamefont {Y.}~\bibnamefont {Levin}},\
  }\bibfield  {title} {\enquote {\bibinfo {title} {Reactive monte carlo
  simulations for charge regulation of colloidal particles},}\ }\href@noop {}
  {\bibfield  {journal} {\bibinfo  {journal} {The Journal of Chemical Physics}\
  }\textbf {\bibinfo {volume} {156}},\ \bibinfo {pages} {014108} (\bibinfo
  {year} {2022}{\natexlab{b}})}\BibitemShut {NoStop}%
\bibitem [{\citenamefont {Guggenheim}(1930)}]{GuggenheimII}%
  \BibitemOpen
  \bibfield  {author} {\bibinfo {author} {\bibfnamefont {E.~A.}\ \bibnamefont
  {Guggenheim}},\ }\bibfield  {title} {\enquote {\bibinfo {title} {On the
  conception of electrical potential difference between two phases. ii},}\
  }\href@noop {} {\bibfield  {journal} {\bibinfo  {journal} {The Journal of
  Physical Chemistry}\ }\textbf {\bibinfo {volume} {34}},\ \bibinfo {pages}
  {1540--1543} (\bibinfo {year} {1930})}\BibitemShut {NoStop}%
\bibitem [{\citenamefont {Craxford}, \citenamefont {Gatty},\ and\ \citenamefont
  {Teorell}(1938)}]{Teorell}%
  \BibitemOpen
  \bibfield  {author} {\bibinfo {author} {\bibfnamefont {S.}~\bibnamefont
  {Craxford}}, \bibinfo {author} {\bibfnamefont {O.}~\bibnamefont {Gatty}}, \
  and\ \bibinfo {author} {\bibfnamefont {T.}~\bibnamefont {Teorell}},\
  }\bibfield  {title} {\enquote {\bibinfo {title} {Xcv. a note on surface
  ph},}\ }\href@noop {} {\bibfield  {journal} {\bibinfo  {journal} {The London,
  Edinburgh, and Dublin Philosophical Magazine and Journal of Science}\
  }\textbf {\bibinfo {volume} {25}},\ \bibinfo {pages} {1061--1066} (\bibinfo
  {year} {1938})}\BibitemShut {NoStop}%
\end{thebibliography}%
				
			\end{document}